\documentclass[5p,times]{elsarticle}
\usepackage{latexsym,amssymb,amsthm}
\usepackage{graphicx}
\AtBeginDocument{\RenewCommandCopy\qty\SI}
\usepackage{physics}
\usepackage[scientific-notation=true]{siunitx}
\usepackage{setspace}
\usepackage{url}
\usepackage{microtype}
\usepackage{tabularx}
\usepackage[utf8]{inputenc}
\usepackage{csquotes}
\usepackage{enumitem}
\usepackage{comment}
\usepackage{adjustbox}
\usepackage{lscape}
\usepackage{array, multirow}
\usepackage{xparse}
\usepackage[dvipsnames]{xcolor}
\usepackage{etoolbox}
\usepackage[section]{placeins}
\usepackage{ragged2e}
\usepackage{tikz}
\usepackage{pgfplots}
\usepackage{supertabular, booktabs}
\usetikzlibrary{trees, positioning, shapes, shadows, arrows.meta}
\usepackage{stfloats}
\usepackage[colorlinks,citecolor=blue,urlcolor=blue,linkcolor=blue]{hyperref}
\usepackage[edges]{forest}
\usepackage{amsmath}

\forestset{
  paper tree/.style={
    for tree={
      grow'=0,
      font=\small,
      edge={draw, semithick},
      child anchor=west,
      parent anchor=south,
      anchor=west,
      calign=fixed edge angles,
      edge path={
        \noexpand\path [draw, \forestoption{edge}]
        (!u.parent anchor) ++(1em,0) |- (.child anchor)\forestoption{edge label};
      },
      before typesetting nodes={
        if n=1 {insert before={[,phantom]}} {}
      },
      fit=band,
      before computing xy={l=1.5em},
    }
  }
}

\definecolor{mygray}{rgb}{0.8,0.8,0.8}

\begin{document}

\begin{frontmatter}

\title{A Survey on Security with Quantum Computing}

\author[cse]{Manik Kumar Sangal}
\ead{manik.sangal22b@iiitg.ac.in}

\author[cse]{Robin Nair}
\ead{robin24b@iiitg.ac.in}

%\author[cse]{Akhirul Islam\corref{cor1}}
\author[trellix]{Akhirul Islam}
\ead{akhirul.islam@trellix.com}

\author[ece]{Sudip Biswas}
\ead{sudip.biswas@iiitg.ac.in}

\author[cse]{Manojit Ghose}
\ead{manojit@iiitg.ac.in}

\affiliation[cse]{organization={Department of Computer Science and Engineering, Indian Institute of Information Technology Guwahati},
                   country={India}}
\affiliation[trellix]{organization={Trellix R\&D, Bengaluru},
                   country={India}}
\affiliation[ece]{organization={Department of Electronics and Communication Engineering, Indian Institute of Information Technology Guwahati}, country={India}}

% \begin{highlights}
% \item Dual-perspective survey: security of quantum systems and quantum-enhanced security
% \item Taxonomy of noise, leakage, compiler, and crosstalk threats in NISQ systems
% \item Thematic analysis of quantum solutions across six cybersecurity domains
% \item Comparison of PQC, QKD, and hybrid quantum-classical defense strategies
% \item Nine concrete open problems identified for quantum security research
% \end{highlights}

\begin{abstract}
Quantum computing has emerged as a transformative computing paradigm capable of solving problems that remain computationally infeasible for classical systems; however, its rapid advancement also introduces significant security, privacy, and reliability concerns. Current noisy intermediate-scale quantum (NISQ) devices are highly susceptible to decoherence, quantum noise, gate instability, measurement errors, and hardware imperfections, which can degrade computational accuracy and potentially expose quantum systems to malicious exploitation. In this context, this survey presents a comprehensive review of security challenges and mitigation strategies associated with quantum computing, focusing on security issues in quantum computers, security threats caused by quantum computers, and security mechanisms developed for quantum systems. The paper examines vulnerabilities in quantum hardware and software, the impact of quantum computing on existing cryptographic infrastructures and cybersecurity mechanisms, and the development of quantum-resilient solutions such as post-quantum cryptography, quantum-safe communication protocols, quantum intrusion detection systems, and quantum-aware software engineering techniques. In addition, the survey discusses emerging applications of quantum technologies in cybersecurity domains including malware detection, network intrusion detection, Internet of Things (IoT) security, and secure communication systems. Furthermore, the paper analyzes existing quantum error mitigation and fault-tolerance approaches designed to improve the robustness and trustworthiness of quantum computation under realistic noisy conditions. By consolidating recent advances, open research challenges, and future directions, this survey provides a structured overview of the evolving intersection between quantum computing and cybersecurity, while serving as a reference for researchers and practitioners working toward secure, resilient, and quantum-ready computing infrastructures.
\end{abstract}

% \begin{abstract}
% Quantum computing is an emerging paradigm that offers solutions to problems beyond the capabilities of classical computing. However, it also introduces new vulnerabilities. Since quantum computing is still in its early stages, current devices are highly susceptible to various forms of noise and error. This article investigates the security implications of quantum systems, focusing on techniques for detecting and mitigating noise and errors that could be exploited to construct malicious quantum circuits. Further, despite such limitations, quantum computing has shown remarkable promise as a powerful and forward-looking field. In this context, the paper also explores the role of quantum computing in cryptography and cybersecurity. It provides a comprehensive survey of the security challenges and opportunities in, by, and for quantum computers, with a focus on challenges and mitigation strategies in noisy intermediate-scale quantum systems.
% \end{abstract}

\begin{keyword}
Quantum computing \sep Security \sep Crosstalk \sep Noise mitigation \sep Post-quantum cryptography \sep Cybersecurity
\end{keyword}

\end{frontmatter}

\section{\textbf{Introduction}} \label{intro}

Quantum computing (QC) has emerged as a transformative computing paradigm capable of solving certain classes of problems that are computationally infeasible for classical systems. By exploiting quantum-mechanical principles such as superposition and entanglement, quantum computers offer significant advantages in optimization, cryptanalysis, simulation, and machine learning (ML), motivating substantial investments from both academia and industry. Major technology companies, including IBM~\cite{steffen2011quantum}, Google~\cite{7934217}, and Microsoft~\cite{Aghaee2025}, have accelerated the development of cloud-accessible quantum platforms, enabling broader experimentation and research in quantum applications. Despite these advancements, quantum systems remain highly susceptible to noise, decoherence, and operational instability, making reliability and security fundamental challenges in practical quantum computing. Quantum hardware is inherently fragile, and even minor environmental disturbances can alter qubit states and compromise computational correctness. Among the prominent concerns, cross-talk~\cite{1,4,17,18,19,20,62,63,64} in shared quantum environments can result in unintended interference between simultaneously executed quantum programs, potentially leading to information leakage and compromised confidentiality. Such vulnerabilities become increasingly critical in multi-tenant cloud-based quantum infrastructures, where multiple users access the same physical hardware resources.

Consequently, securing quantum systems has become a major research focus~\cite{8,11,12,24}. Existing studies investigate mitigation strategies at different abstraction levels, ranging from architectural isolation and workload partitioning~\cite{9,10} to advanced quantum error correction and noise-aware compilation techniques. Partitioning approaches attempt to isolate user workloads and reduce interference among concurrent quantum executions, although they often introduce additional system overhead and reduced hardware utilization. Similarly, quantum error correction mechanisms aim to identify and mitigate errors caused by decoherence, gate imperfections, and cross-talk, thereby improving both reliability and security. At the same time, the rapid progress of quantum computing introduces significant implications for classical cybersecurity. Many widely deployed cryptographic schemes, particularly public-key systems such as RSA and ECC, rely on mathematical assumptions that can be efficiently broken by sufficiently powerful quantum computers~\cite{Mavroeidis_2018,26,27,29,30,nist2016pqcrypto,10.1007/978-3-642-40084-1_21}. This emerging threat has accelerated research into post-quantum cryptography (PQC)~\cite{27,29}, which seeks to develop cryptographic algorithms resistant to quantum attacks while remaining deployable on classical hardware. In parallel, quantum key distribution (QKD) has gained attention as a quantum-native approach for secure communication, leveraging the principles of quantum mechanics to detect eavesdropping attempts and ensure communication integrity.

\begin{table*}[t]
\centering
\caption{Summary of survey references and their scope}
\label{tabExistingSurvey}
\begin{adjustbox}{max width=\textwidth}
\begin{tabular}{p{1.5cm}p{1.2cm}p{7.8cm}p{5.2cm}}
\hline
\textbf{Ref.} & \textbf{Year} & \textbf{Brief description} & \textbf{Coverage scope} \\ \hline
\cite{survey-10460214} & 2024 &  Tutorial and survey on vulnerabilities in quantum computing systems & QC vulnerabilities \\ %\hline
\cite{survey-8932459} & 2020 & Post-quantum IoT systems and challenges & PQ IoT \\ %\hline
\cite{survey-10401941} & 2024  & Post-quantum blockchains for IoT security & PQ blockchain, IoT \\ %\hline
\cite{survey-latticebased} & 2024 &  Lattice-based cryptosystems for smart-grid security & Lattice crypto, smart grids \\ %\hline
\cite{survey-10807207} & 2024 & Quantum resilience of symmetric cryptographic primitives & Symmetric crypto \\ %\hline
\cite{survey-GLISIC2024100039} & 2024 &  Quantum and post-quantum cryptography for wireless/6G networks & QC/PQC, wireless \\ %\hline
\cite{survey-WICAKSANA2025100752} & 2025  & Quantum-safe solutions for blockchain security & PQ blockchain \\ %\hline
\cite{survey-securityvulnerabilitiesquantumcloud} & 2025 &  Security challenges in multi-tenant quantum cloud & QC cloud security \\ %\hline
\cite{survey-pqba} & 2025 &  Post-quantum threats to edge devices and systems & PQ edge security \\ %\hline
\cite{survey-11006701} & 2025 &  Lattice-based PQC on software and hardware platforms & Lattice PQC \\ \addlinespace
\textbf{This work} & \textbf{2026} &  \textbf{Comprehensive survey of security of and by quantum computing} & \textbf{QC noise, crosstalk, compilers, info leakage, PQC, QKD, blockchain, malware, intrusion detection system (IDS), IoT, cybersecurity} \\ \hline
\end{tabular}
\end{adjustbox}
\end{table*}

Beyond cryptography, quantum computing is expected to influence a broad spectrum of cybersecurity domains, including blockchain security~\cite{68-blockchain, 73-blockchain}, intrusion detection systems~\cite{81-intrusion, 82-intrusion, 84-intrusion}, Internet of Things (IoT) security~\cite{38}, software verification~\cite{36}, and secure communication architectures~\cite{85-cyber}. At the same time, quantum systems themselves introduce new attack surfaces related to quantum hardware, software stacks, compilers, and cloud orchestration frameworks. These developments create a dual research challenge: securing quantum computers against emerging threats while simultaneously leveraging quantum technologies to strengthen future cybersecurity systems.

Although several survey articles have explored individual aspects of quantum computing and security, most existing works primarily focus either on quantum cryptography or on isolated security concerns within quantum systems~\cite{47, survey-10460214, survey-8932459, survey-10401941}. A comparative summary of representative existing surveys and the scope of this work is presented in Table~\ref{tabExistingSurvey}. In contrast, this survey adopts a broader and more integrated perspective by examining both the security challenges associated with quantum computing systems and the application of quantum computing to cybersecurity domains. To provide a comprehensive analysis, the surveyed literature was selected based on three primary criteria: (i) relevance to qubit computation, quantum data encoding, and quantum architectures, (ii) investigation of security implications arising from interactions between quantum hardware and software layers, including noise, cross-talk, and information leakage, and (iii) domain-specific studies addressing the impact of quantum computing on cybersecurity applications such as cryptography, blockchain, intrusion detection, software security, and IoT systems. Through this survey, we aim to consolidate recent advances, identify emerging research directions, and highlight open challenges toward the development of secure, reliable, and scalable quantum computing ecosystems that can coexist with classical infrastructures.

\begin{figure}[!htb]
\centering
\begin{forest}
  for tree={
    folder,
    grow'=0,
    font=\small,
    s sep=0.8ex,
    l sep=1em,
  }
  [, phantom
    [{\textbf{\pmb{\S} \ref{intro} Introduction}}
        %[\ref{scope-paper-selection} Scope and paper selection]
    ]
    [{\textbf{\pmb{\S} \ref{S1} Background}}]
    [{\textbf{\pmb{\S} \ref{S2} Security challenges and solutions in quantum computing}}
        [\ref{sec:noise} Noise in quantum systems: challenges and mitigation]
        [\ref{sec:infoleakage} Information leakage and countermeasures]
        [\ref{sec:compilers} Third-party compiler threats and defenses]
        [\ref{sec:crosstalk} Crosstalk: characterization and mitigation]
    ]
    [{\textbf{\pmb{\S} \ref{S4} Quantum computing for cybersecurity applications}}
        [\ref{sec:crypto} Cryptography and post-quantum cryptography]
        [\ref{sec:blockchain} Blockchain security]
         [\ref{sec:malware} Malware detection]
         [\ref{sec:itsecurity} {IT, software, and network intrusion detection}]
         [\ref{sec:iot} IoT security]
    ]
    [{\textbf{\pmb{\S} \ref{benchmark} Benchmarks used}}
    ]
    [{\textbf{\pmb{\S} \ref{sec:future} Open problems and future directions}}
    ]
    [{\textbf{\pmb{\S} \ref{conclusion} Conclusions}}]
    ]
  ]
\end{forest}

\caption{Organization of the paper.}
\label{fig:sections}
\end{figure}

The remainder of this paper is organized as in Figure~\ref{fig:sections}. Section~\ref{S1} provides a brief overview of quantum computing. Section~\ref{S2} addresses security challenges in quantum systems and their corresponding mitigation strategies, organized by topic. Section~\ref{S4} examines how quantum computing can enhance cybersecurity across multiple domains, with each subsection presenting both the challenges and proposed solutions. Section~\ref{benchmark} summarizes the simulators and benchmarks used across the surveyed studies. Section~\ref{sec:future} identifies open problems and future research directions, and Section~\ref{conclusion} concludes the paper.

\begin{table*}[!tb]
\footnotesize
\centering
\caption{Threat model for vulnerabilities in quantum computing systems (Section~\ref{S2})}
\label{tableSurvey1}
\renewcommand{\arraystretch}{1}

\begin{adjustbox}{width=\textwidth}
\begin{tabular}{%
p{3cm}
p{4.2cm}
p{4.2cm}
p{4.2cm}
p{1cm}
}
\toprule
\textbf{Threat category} & \textbf{Vector / Cause} &
\textbf{Security impact} & \textbf{Exploitation conditions} & \textbf{Section} \\
\midrule

Quantum noise exploitation &
Gate imperfections, decoherence, readout errors, and crosstalk &
Incorrect computation results and reduced computational fidelity &
Execution on noisy quantum hardware without full error correction &
\S\ref{sec:noise} \\ \addlinespace

Noise-induced information leakage &
Residual quantum states and systematic noise across executions &
Partial disclosure of quantum state information or output bias &
Repeated circuit execution under persistent noise conditions &
\S\ref{sec:infoleakage} \\ \addlinespace

Malicious circuit manipulation &
Adversarial circuit designs exploiting noise-induced effects &
Integrity violations and unintended information exposure &
Ability to influence or design executed quantum circuits &
\S\ref{sec:infoleakage} \\ \addlinespace

Quantum side-channel attacks &
Timing behavior, energy usage, and micro-architectural signals &
Inference of circuit behavior or sensitive computation patterns &
Availability of observable side-channel information &
\S\ref{sec:infoleakage} \\ \addlinespace

Untrusted Third-Party Compilation &
External compiler access to full quantum circuit descriptions &
Intellectual property leakage and circuit confidentiality loss &
Reliance on untrusted or opaque compilation services &
\S\ref{sec:compilers} \\ \addlinespace

Compiler-based Trojan insertion &
Malicious gate reordering, qubit mismapping, or hidden subcircuits &
Biased computation outcomes and undetected result manipulation &
Limited verification due to probabilistic quantum execution &
\S\ref{sec:compilers} \\ \addlinespace

Platform-specific compilation abuse &
Hardware-aware over optimization (e.g., SWAP insertion, shuttling) &
Increased noise levels and reduced gate fidelity &
Dependence on specific hardware constraints and layouts &
\S\ref{sec:compilers} \\ \addlinespace

Multi-qubit crosstalk &
Unintended qubit--qubit interactions (e.g., ZZ coupling) &
Correlated errors, data leakage, and degraded logical operations &
Simultaneous or closely scheduled multi-qubit operations &
\S\ref{sec:crosstalk} \\ \addlinespace

Crosstalk in multi-programming regimes &
Concurrent execution of multiple user circuits &
Reduced correctness probability and output interference &
Shared quantum hardware in multi-tenant environments &
\S\ref{sec:crosstalk} \\

\bottomrule
\end{tabular}
\end{adjustbox}
\end{table*}

\section{Background}\label{S1}

Quantum computing is an emerging computational paradigm that leverages principles of quantum mechanics, such as superposition and entanglement, to achieve computational advantages over classical computing for certain problem classes~\cite{10.1145/367701.367709,10.1093/nsr/nwy072}. Unlike classical computers, which operate on binary bits, quantum computers use quantum bits (qubits) as the fundamental unit of information. While a classical bit can exist only in one of two states, $0$ or $1$, a qubit can exist in a superposition of both basis states simultaneously, represented as in Equation~(\ref{qubits}):

\begin{equation}
    \label{qubits}
    |\psi\rangle = \alpha |0\rangle + \beta |1\rangle
\end{equation}

where $\alpha$ and $\beta$ are complex-valued probability amplitudes satisfying the normalization condition as in Eqaution~(\ref{norm-cond}):

\begin{equation}
    \label{norm-cond}
    |\alpha|^2 + |\beta|^2 = 1
\end{equation}

This superposition property enables quantum systems to represent multiple computational states simultaneously, forming the basis for quantum parallelism and allowing potential speedups for problems such as factorization and optimization.

Another fundamental quantum property is entanglement, in which the states of multiple qubits become correlated such that the joint quantum state cannot be described independently of its constituent qubits, regardless of their physical separation. Formally, a multi-qubit state is considered entangled if it cannot be expressed as the tensor product of individual qubit states:

\begin{equation}
    \label{mqbit}
    |\psi\rangle = |\psi_1\rangle \otimes |\psi_2\rangle
\end{equation}

Entanglement enables non-classical correlations between qubits and plays a critical role in quantum algorithm design, quantum error correction, and communication protocols such as quantum teleportation and quantum key distribution~\cite{RevModPhys.81.865,PhysRevA.65.012101}.

Computations in quantum systems are represented using quantum circuits composed of quantum gates. Unlike classical logic gates, quantum gates are reversible and mathematically represented by unitary matrices. By exploiting superposition and entanglement, quantum circuits can perform computations that offer significant advantages for certain applications. Notable examples include Shor’s algorithm for integer factorization and Grover’s algorithm for searching unsorted databases, both of which provide substantial speedups over their best-known classical counterparts~\cite{Mosca2009}.

%% ================================================================
%% SECTION III: SECURITY CHALLENGES AND SOLUTIONS IN QC
%% ================================================================
\section{Security challenges and solutions in quantum computing} \label{S2}

This section discusses the key vulnerabilities in quantum computing systems and their corresponding mitigation strategies, organized by topic. Table~\ref{tableSurvey1} provides a comprehensive threat model cataloging all vulnerabilities in quantum computing systems discussed in this section.

\subsection{Noise in quantum systems: challenges and mitigation}\label{sec:noise}

\subsubsection{The challenge}

Noise is one of the most persistent challenges in quantum computing due to the inherent fragility of quantum systems. Key error types in quantum systems include gate errors caused by imperfect quantum operations and decoherence resulting from the loss of quantum states through interactions with the surrounding environment. Other major sources of errors include readout errors arising from inaccuracies during measurement and crosstalk caused by unintended interactions between qubits~\cite{xu2023classification}. These errors generally manifest as either coherent noise, which originates from predictable sources such as imperfect gate implementations or environmental coupling, or incoherent noise, which results from unpredictable effects including decoherence, thermal fluctuations, and spontaneous emission. Such noise-induced errors may be exploited to design malicious circuits, posing significant security risks to quantum computations. Noise may also arise from adversarial hardware manipulation, including supply-chain tampering, where malicious modifications introduced during fabrication or calibration embed systematic errors into quantum devices.

\subsubsection{Mitigation strategies}
To address these challenges, many innovative noise mitigation techniques have been developed and classified into high-level, low-level, and bridge-level categories, as proposed by Guimar\~{a}es \textit{et al.}~\cite{2}. Table~\ref{tab:mitigation-summary} summarizes the trade-offs, advantages, and limitations of these techniques. We also illustrate all techniques covered in this section in Figure~\ref{figQEMT} for better presentation.

\setlength{\tabcolsep}{4pt}
\setlength{\emergencystretch}{3em}

\begin{table*}[!tb]
\centering
\caption{Comparative summary of quantum error mitigation techniques}
\label{tab:mitigation-summary}
\scriptsize
\renewcommand{\arraystretch}{1.0}
\setlength{\tabcolsep}{3pt}
\begin{tabularx}{\textwidth}{@{}lllXllX@{}}
\toprule
\textbf{Technique} & \textbf{Level} & \textbf{Noise type} & \textbf{Key principle} & \textbf{Overhead} & \textbf{HW Access} & \textbf{Advantages} \\ \addlinespace
\midrule
ZNE & High & Incoherent & Runs circuits at varying noise levels; extrapolates to zero noise & Mod--High & Low & Simple; works on current devices \\ \addlinespace
QPM & High & Coh./Incoh. & Reconstructs ideal expectation values via probabilistic reweighting & High & Moderate & Broad applicability \\ \addlinespace
COMDAP & High & Coherent & Integrates composite pulses and dynamical decoupling & Low--Mod & High & Enhances gate robustness \\ \addlinespace
Rand.\ compiling & Bridge & Coherent & Randomizes gate sequences; converts coherent noise to stochastic & Low--Mod & Moderate & Robust without calibration \\ \addlinespace
Symmetry verif. & Bridge & Incoherent & Filters results inconsistent with conserved system symmetries & Low & Low & Accuracy gain when symmetries exist \\ \addlinespace
MEM & Bridge & Incoherent & Calibrates and corrects readout errors using confusion matrices & Moderate & Low & Easy to implement \\ \addlinespace
QUIET & Bridge & Coh./Incoh. & Quasi-probabilistic inference and tensor networks & High & Moderate & Scalable to large systems \\ \addlinespace
DAEM & Bridge & Coh./Incoh. & Learns adaptive noise models during training & Moderate & Moderate & Adaptive; data-driven \\ \addlinespace
ANN-QEM & Bridge & Coh./Incoh. & Neural networks predict and correct noise-induced deviations & Moderate & Moderate & ML-based flexibility \\ \addlinespace
Hybrid QC-EM & Bridge & Coh./Incoh. & Classical post-processing with quantum denoising subroutines & Moderate & Moderate & Balances QC and classical resources \\ \addlinespace
Dyn.\ decoupling & Low & Coherent & Inserts control pulses to average out environmental noise & Low & High & Hardware-effective \\
\bottomrule
\end{tabularx}
\end{table*}

%\textbf{High Level Techniques:}
\section*{High Level Techniques}
 High-level techniques do not require detailed knowledge of the underlying physical hardware implementation. They focus primarily on signal reconstruction and statistical methods to mitigate noise without needing direct control over the hardware. These techniques include:
\begin{enumerate}

\item \textbf{Zero-noise extrapolation (ZNE)~\cite{giurgica2020digital, he2020zero}:} ZNE is a widely used quantum error mitigation technique that improves the accuracy of noisy quantum computations without requiring additional quantum resources. It operates by intentionally executing the same quantum program under multiple amplified noise levels and then extrapolating the measured results to estimate the outcome in an ideal noiseless setting. The method typically involves two stages: first, scaling the effective noise of the quantum hardware and collecting outputs at several noise levels; second, applying extrapolation techniques to infer the error-free result. Because ZNE can be implemented on existing hardware without modifying quantum circuits significantly, it has become a practical and popular approach for mitigating noise in near-term quantum devices.

\item \textbf{Quasi-probability method~\cite{2, hofer2017quasi, ferrie2011quasi}:} 
Quasi-probability methods, derived from phase-space representations of quantum theory, mitigate quantum noise by expressing noisy operations as probabilistic combinations of ideal quantum operations using probability-like distributions that may take negative or complex values. This framework bridges classical and quantum mechanics by enabling expectation-value estimation while capturing non-classical quantum behavior. Although quasi-probability techniques can mitigate both coherent and incoherent noise effectively, their practical applicability is limited by exponential sampling overhead that grows with circuit depth, restricting their use primarily to shallow circuits on NISQ devices.

% Models noisy operations as probabilistic mixtures of ideal operations:
% \begin{equation}
%     \mathcal{E}(\rho) = \sum_i p_i \mathcal{U}_i(\rho)
% \end{equation}
% where \(\mathcal{E}\) is the noisy operation, \(\mathcal{U}_i\) are unitary operations, and \(p_i\) are probabilities. Despite their ability to mitigate both coherent and incoherent noise, quasi-probability methods are constrained by exponential sampling overhead proportional to circuit depth. This scaling limit restricts their utility to shallow circuits and prevents practical deployment on long-depth noisy intermediate-scale quantum (NISQ) devices.

\begin{figure}[!tb]
\centering
\resizebox{\columnwidth}{!}{%
\begin{tikzpicture}[
    node distance=1.0cm and 1.2cm,
    every node/.style={font=\scriptsize},
    box/.style={draw, rounded corners, fill=blue!10, minimum width=2.3cm, minimum height=0.7cm, text width=2.5cm, align=center},
    level/.style={draw, rounded corners, fill=gray!20, minimum width=1.8cm, minimum height=0.7cm, align=center}
]
\node[level] (root) {Error Mitigation\\Techniques};
\node[level, right=of root,yshift=4.2cm] (high) {High Level};
\node[level, right=of root] (bridge) {Bridge Level};
\node[level, right=of root,yshift=-4.2cm] (low) {Low Level};
\node[box, right=of high, yshift=0.6cm] (zne) {ZNE \cite{zhang2025demonstrating}};
\node[box, right=of high, yshift=-0.6cm] (qpm) {QPM \cite{2}};
\node[box, right=of bridge, yshift=2.7cm] (rc) {Rand.\ Compiling \cite{2}};
\node[box, right=of bridge, yshift=1.8cm] (sv) {Symmetry Verif.\ \cite{2}};
\node[box, right=of bridge, yshift=0.9cm] (me) {MEM \cite{2}};
\node[box, right=of bridge] (quiet) {QUIET \cite{muqeet2024quiet}};
\node[box, right=of bridge, yshift=-0.9cm] (daem) {DAEM/DAEN \cite{3}};
\node[box, right=of bridge, yshift=-1.8cm] (ann) {ANN-QEM \cite{adeniyi2025}};
\node[box, right=of bridge, yshift=-2.7cm] (hqcem) {Hybrid QC-EM \cite{14}};
\node[box, right=of low, yshift=0.5cm] (comdap) {COMDAP \cite{23}};
\node[box, right=of low, yshift=-0.5cm] (dd) {Dyn.\ Decoupling \cite{biercuk2009optimized}};
\draw[-{Latex}] (root.east) -- (high.west);
\draw[-{Latex}] (root.east) -- (bridge.west);
\draw[-{Latex}] (root.east) -- (low.west);
\draw[-{Latex}] (high.east) -- (zne.west);
\draw[-{Latex}] (high.east) -- (qpm.west);
\draw[-{Latex}] (bridge.east) -- (rc.west);
\draw[-{Latex}] (bridge.east) -- (sv.west);
\draw[-{Latex}] (bridge.east) -- (me.west);
\draw[-{Latex}] (bridge.east) -- (quiet.west);
\draw[-{Latex}] (bridge.east) -- (daem.west);
\draw[-{Latex}] (bridge.east) -- (ann.west);
\draw[-{Latex}] (bridge.east) -- (hqcem.west);
\draw[-{Latex}] (low.east) -- (comdap.west);
\draw[-{Latex}] (low.east) -- (dd.west);
\end{tikzpicture}%
}
\caption{Types of quantum error mitigation techniques}
\label{figQEMT}
\end{figure}
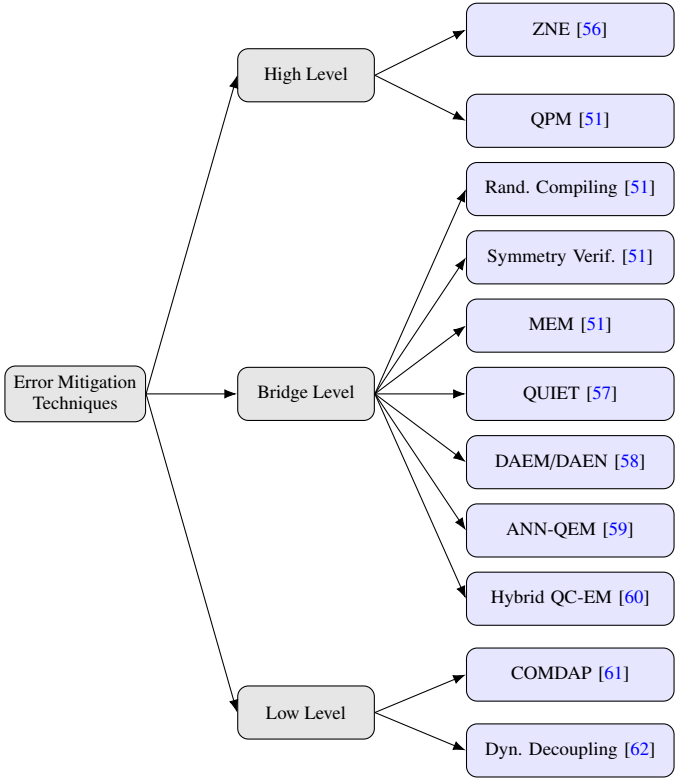

\item \textbf{COMDAP: Community-based dynamic allocation partitioning~\cite{6, 23}}: COMDAP is a system-level architectural framework designed to enhance error resilience in multi-programmed quantum environments. Distinguishing itself from physical-layer noise suppression techniques, COMDAP mitigates inter-program crosstalk through strategic resource partitioning. The framework utilizes the Louvain algorithm to identify densely connected qubit communities within the hardware topology, which are then evaluated using a Connectivity and Reliability Index (CRI). This index aggregates topological features, including path length and node connectivity, with hardware error rates to prioritize high-fidelity clusters for concurrent execution. While COMDAP optimizes throughput in multi-tenant scenarios, its efficacy is contingent upon accurate hardware characterization, and it does not provide direct mitigation for stochastic errors such as decoherence.
\end{enumerate}

\section*{Bridge techniques}
These methods strike a balance between hardware abstraction and circuit-specific optimization. They typically operate at the compiler or circuit transformation level, requiring moderate access to quantum system details. Some examples of these techniques.

\begin{enumerate}
\item \textbf{Randomized compiling (RC)~\cite{hashim2020randomized, 2}:} 
Randomized compiling is a protocol designed to mitigate coherent errors in quantum algorithms by transforming them into stochastic noise channels, thereby reducing unpredictable error accumulation and enabling more accurate prediction of algorithmic performance from error rates measured via cycle benchmarking. Compared to approaches such as Pauli frame randomization and simple Pauli twirling, RC is more scalable and generalizable, and does not require prior knowledge of the underlying error model. It achieves this by inserting random single-qubit virtual twirling gates into a circuit and recompiling the circuit such that the overall unitary operation is preserved. As a result, RC generates multiple logically equivalent randomized versions of the original circuit without increasing circuit depth, and averaging over these circuit realizations effectively tailors coherent errors into stochastic noise.

% This approach transforms coherent noise into stochastic Pauli noise by randomly inserting and removing gates during compilation.

% \begin{equation}
%     \mathcal{C}_{\text{RC}}(\rho) = \frac{1}{N} \sum_{i=1}^N \mathcal{U}_i \mathcal{C} \mathcal{U}_i^\dagger (\rho)
% \end{equation}

% where \(\mathcal{C}\) is the circuit, \(\mathcal{U}_i\) are random unitary operations, and \(N\) is the number of randomizations~\cite{2}.

\item \textbf{Symmetry verification~\cite{bonet2018low, 2}:}
Symmetry verification is a quantum error mitigation technique that improves the accuracy of quantum computations by detecting and filtering quantum states that violate known physical symmetries, such as parity or particle-number conservation. Acting as a stabilizer-based filter, it identifies instances where noise drives the system outside the valid physical subspace and rejects or post-processes these unphysical results. This approach is commonly used to enhance the performance of near-term quantum algorithms, including the variational quantum eigensolver (VQE) and quantum qpproximate optimization algorithm. Symmetry verification can be implemented through ancilla-assisted measurements, in-line Pauli symmetry checks, or classical post-processing, and has proven particularly effective in mitigating qubit relaxation and excitation errors in molecular simulations and lattice gauge theory applications.

\item \textbf{Measurement error mitigation (MEM)~\cite{ 2, bravyi2021mitigating, funcke2022measurement}:} 
Measurement error mitigation consists of classical post-processing techniques used to reduce readout errors in quantum computers, acting as a bridge until full error correction is feasible. It works by calibrating against known hardware noise and applying inverse operations to improve the accuracy of measured data, often utilizing methods like matrix-free measurement mitigation or twirled readout error extinction.

\item \textbf{Quantum information error tracer (QUIET)~\cite{janardan2016analytical, muqeet2024quiet}}: QUIET is a bridge-level quantum noise mitigation tool proposed by Muqeet \textit{et al.}~\cite{muqeet2024quiet, janardan2016analytical} designed to improve the reliability of sampling-based outputs of quantum software by implementing the  quantum learning with error analysis and reduction (QLEAR) technique. It works independently of specific quantum execution frameworks and is compatible with various quantum devices. Its performance was evaluated using the Hellinger distance to measure the closeness between the ideal, noisy, and mitigated outputs.

\item \textbf{Neural network-based error mitigation}~\cite{3, ponsi2023mitigation}: A promising approach to mitigating quantum noise is the data augmentation empowered neural model (DAEM), proposed by Liao \textit{et al.}~\cite{3, ponsi2023mitigation}. This quantum-aware neural network framework is designed to reduce computational errors caused by noisy environments. DAEM achieves this by leveraging the power of neural networks to correct output distortions in quantum computations.
%~\cite{adeniyi2025}

\item \textbf{Adaptive neural network-based quantum error mitigation (ANN-QEM)}~\cite{adeniyi2025}: A ML-driven, bridge-level framework that employs adaptive neural networks for real-time error identification and correction. Unlike static noise models, ANN-QEM dynamically adjusts to specific circuit characteristics through two modules: (1) a classifier for error detection and (2) a regression module for probability compensation. 

The framework's methodology follows a structured pipeline:
\begin{enumerate}
    \item \textbf{Data generation:} Execution of diverse quantum circuits with varying depths to capture broad quantum behaviors.
    \item \textbf{Feature extraction:} Characterization of different noise sources, including Pauli noise, depolarizing noise, and amplitude damping noise, using quantum circuit configurations and statistical features derived from measurement outcomes.
\end{enumerate}

Despite its adaptability, ANN-QEM faces significant overhead in acquiring high-fidelity training data, risking a \enquote{circularity problem} where clean data is required to train models meant to generate clean data. Furthermore, neural network corrections may yield correct outputs from fundamentally flawed quantum states, potentially introducing false confidence in the underlying state reliability.

\item \textbf{Hybrid quantum-classical error mitigation}~\cite{14, ponsi2023mitigation, popirlan2023hybrid}: Combines quantum error detection with classical correction, offering a favorable balance between effectiveness and overhead at the cost of additional post-processing latency. While hybrid approaches offer a favorable balance between mitigation effectiveness and resource overhead, they typically introduce additional classical post-processing costs and latency, which may limit their use in time-sensitive quantum workloads.
\end{enumerate}

\section*{Low-level techniques}
 Hardware-aware methods operate at the physical layer by directly manipulating qubit dynamics and gate sequences to suppress noise. These strategies necessitate precise characterization of the underlying architecture, including qubit topology, decoherence rates, and cross-resonance effects, to optimize pulse-level control and timing.

Dynamical decoupling (DD) serves as a foundational physical-layer technique, often integrated into more complex error mitigation strategies~\cite{paz2013optimally}. Inspired by spin-echo sequences in nuclear magnetic resonance, DD suppresses decoherence by applying periodic, precisely timed control pulses, typically $\pi$-pulses to idle qubits. These qubit-flip operations effectively time-average environmental interactions to zero, significantly mitigating low-frequency or static noise. As noted by Biercuk \textit{et al.} \cite{biercuk2009optimized}, the frequent inversion of the qubit state preserves quantum information and extends coherence times without the substantial resource overhead required for active quantum error correction~\cite{khadirsharbiyani2024minimizing}.

% \begin{enumerate}
% \item \textbf{Dynamical decoupling (DD):} Inserts precisely timed $\pi$-pulses during idle periods to average out environmental noise, extending coherence times without active error correction~\cite{biercuk2009optimized}. Analogous to spin echo techniques in NMR, DD is effective against slowly varying or static noise.
% \end{enumerate}
%We note that this three-level taxonomy is a simplification; many techniques (e.g., DD when implemented at the compiler level) span multiple categories.

\subsubsection{Synthesis}
Among these mitigation strategies, ZNE and QPM offer the broadest applicability without hardware modification, but suffer from sampling overhead that grows with circuit depth. COMDAP and RC provide system-level improvements suited to multi-tenant environments, while QUIET and ANN-QEM leverage ML for adaptive correction at the cost of training data requirements. DD remains the most hardware-effective technique for coherent noise but requires precise pulse control. The choice among these approaches depends on the target noise model, available hardware access, and acceptable overhead, suggesting that practical deployments will likely combine multiple techniques across abstraction levels.

\subsection{Information leakage and countermeasures}
\label{sec:infoleakage}
\subsubsection{The challenge}
Quantum noise introduces significant security challenges in quantum computing by enabling unintended information leakage during computation~\cite{11}. Noise-induced disturbances can expose partial information from quantum operations, reducing both the confidentiality and reliability of computation outputs. In addition, adversaries may exploit these errors to construct malicious quantum circuits that function as covert channels for leaking sensitive information. A particularly important concern is state leakage, where residual quantum states persist across repeated circuit executions, allowing attackers to infer information over multiple runs through systematic error analysis~\cite{11}. 

Quantum systems are also vulnerable to side-channel attacks similar to those observed in classical computing. Such attacks exploit micro-architectural characteristics, including execution timing and energy consumption patterns, to reconstruct circuit behavior and extract confidential information~\cite{12,24}. To mitigate noise and improve computation fidelity, some quantum software ecosystems rely on third-party compilers that optimize gate operations and reduce execution overhead~\cite{10}. However, dependence on external compilation frameworks introduces additional risks, including potential circuit manipulation and data leakage during the optimization process.
%\subsubsection{The challenge}
%Quantum noise can compromise the confidentiality of quantum computations through two primary mechanisms~\cite{11}: (1)~partial data exposure during noisy operations, and (2)~adversarial exploitation of noise to design malicious circuits that act as covert channels. A key concern is state leakage, when circuits are executed repeatedly, residual quantum states persist between runs, enabling information extraction across iterations~\cite{11}. Side-channel attacks further threaten quantum systems by exploiting timing or energy consumption patterns to reconstruct circuit behavior~\cite{12,24}. In response, some ecosystems have adopted third-party compilers claiming faster processing and enhanced fidelity~\cite{10}; however, reliance on external compilers introduces additional risks discussed in Section~\ref{sec:compilers}.

\subsubsection{Countermeasures}
To address information leakage and side-channel threats in quantum computing, researchers have proposed several defense mechanisms spanning encryption-based protection, circuit obfuscation, and side-channel mitigation~\cite{goertzel2025efficient,upadhyay2025quantum,john2025quantum,das2025impact,kumar2025context}. One important approach is the use of quantum one-time pads (QOTPs) proposed by Xu \textit{et al.}~\cite{11}, which extend the classical one-time pad concept to quantum systems by protecting both the amplitude and phase information of qubits through Pauli-X and Pauli-Z operations. QOTPs provide confidentiality for quantum data during computation and communication. For outsourced quantum computation on untrusted servers, Goertzel \textit{et al.}~\cite{goertzel2025efficient} propose a quantum-secure homomorphic encryption framework that combines module learning with errors (MLWE) and bounded natural super-functors, enabling quantum programs to operate directly on encrypted data without revealing sensitive information to the server.

Another major class of defenses focuses on quantum circuit obfuscation to protect intellectual property and prevent circuit reconstruction attacks. Suresh \textit{et al.}~\cite{9} propose inserting dummy controlled-NOT (CNOT) gates into quantum circuits to intentionally corrupt outputs and make reverse engineering difficult. The effectiveness of this obfuscation is evaluated using total variation distance (TVD), which measures the deviation between original and obfuscated outputs. However, adding extra gates increases decoherence and noise, creating a trade-off between security and circuit fidelity. Building on this idea, Rehman \textit{et al.}~\cite{rehman2025opaque} introduce OPAQUE, which leverages phase-gate rotation angles for obfuscation instead of inserting additional key qubits and controlled gates, thereby reducing physical overhead. Similarly, Bartake \textit{et al.}~\cite{bartake2025obfusqate} propose the ObfusQate framework, which incorporates multiple circuit-level and control-flow-level obfuscation techniques, including inverse gates, composite gates, cloaked gates, and delayed gates, to make reverse engineering significantly harder while preserving circuit functionality.

To counter power side-channel attacks, Ferhat Erata \textit{et al.}~\cite{12} investigate per-channel single-trace analysis, where attackers use individual qubit power traces to reconstruct quantum circuits. They formulate the reconstruction process as a mixed integer linear programming problem that models circuit components and power-consumption behavior. Beyond individual defense mechanisms, broader quantum cybersecurity efforts~\cite{13} are also underway to establish taxonomies of quantum-specific vulnerabilities and develop dedicated testbed environments for evaluating the security and resilience of quantum infrastructures. Together, these countermeasures highlight the growing effort toward building secure, scalable, and resilient quantum computing systems while balancing security guarantees against noise, decoherence, and implementation overhead.

\subsection{Third-party compiler threats and defenses}\label{sec:compilers}

\subsubsection{The challenge}

The growing reliance on third-party quantum compilers has introduced significant security and reliability challenges, particularly when the compiler cannot be fully trusted. Untrusted compilers may generate erroneous outputs due to transformation bugs, inefficient optimizations, or malicious modifications that increase circuit noise and degrade performance. One of the most serious threats is intellectual property theft, where adversaries reverse-engineer quantum circuits to extract proprietary algorithms or sensitive information~\cite{53}. In addition, malicious compilers may perform qubit mismapping or insert Trojan circuits that subtly alter gate layouts to bias computation outcomes without being easily detected~\cite{55}. Since quantum computations are inherently probabilistic, these manipulations can evade traditional verification mechanisms, making dishonest modifications difficult to identify~\cite{54}. 

The impact of such threats varies across quantum hardware platforms. In superconducting architectures, vulnerabilities often arise from the heavy dependence on compiler optimizations to minimize SWAP operations caused by limited qubit connectivity. Similarly, in ion-trap systems, untrusted compilers may introduce unnecessary ion shuttling or inefficient trap mappings, leading to increased execution overhead and reduced gate fidelity. These challenges highlight the need for secure and trustworthy compilation frameworks capable of preserving both circuit integrity and computational correctness across diverse quantum architectures.

\subsubsection{Defenses}
To mitigate the risks introduced by untrusted third-party quantum compilers, several security mechanisms have been proposed~\cite{wang2025tetrislock,rehman2025opaque,bartake2025obfusqate}. One important direction is \textit{split compilation}, proposed by Saki \textit{et al.}~\cite{10}, where a quantum circuit is divided into multiple sub-circuits and their execution order is obfuscated so that no single compiler can reconstruct the entire circuit design. A related approach proposed by Das \textit{et al.}~\cite{53} inserts reversible random circuits into the original quantum program to intentionally corrupt the circuit structure during compilation. After compilation, the inverse of the inserted circuit is appended to restore the original functionality. This technique helps protect the intellectual property of the circuit by making reverse engineering significantly more difficult.

To improve reliability in environments containing both trusted and untrusted hardware, Upadhyay \textit{et al.}~\cite{54} propose an adaptive shot distribution strategy that distributes repeated circuit executions, known as quantum shots, across multiple hardware devices. Their framework dynamically allocates more shots to hardware platforms identified as reliable at runtime, reducing the impact of tampered or low-quality devices. More broadly, recent compiler-centric defenses focus on protecting circuits before they leave the trusted client environment. These approaches include dynamic compilation strategies, circuit masking, and architecture-level randomness designed to prevent adversaries from inferring circuit structure, functionality, or encoding information during compilation.

As quantum ML and cloud-based quantum services continue to expand, researchers have also proposed defenses specifically targeting quantum neural networks (QNNs). Kundu \textit{et al.}~\cite{55} introduce safeguarding training and inferencing of QNNs, where two QNNs are trained in parallel to generate obfuscated intermediate outputs that are later aggregated to obtain the final result. In the context of quantum machine learning as a service, Upadhyay \textit{et al.}~\cite{65} propose two protection mechanisms: masking labels through secret qubit measurement combinations known only to the user, and modified cost functions that penalize incorrect qubit measurements to mislead adversaries attempting to infer labels.

Another recent framework, called \textit{TetrisLock}, is proposed by Wang \textit{et al.}~\cite{wang2025tetrislock}. This method inserts reversible random circuits and their corresponding inverse operations into the original circuit while preserving the overall computation. The transformed circuit is then partitioned into interdependent sub-circuits using an interlocking structure such that individual segments reveal minimal information about the original design. This significantly increases the difficulty of circuit reconstruction, particularly in scenarios involving colluding compilers. Similarly, Upadhyay \textit{et al.}~\cite{upadhyay2025quantum} propose a hybrid quantum-classical encoding obfuscation framework that analyzes structural patterns in parameterized quantum circuits to identify encoding schemes such as angle, amplitude, and basis encoding, highlighting the growing risk of encoding-identification attacks in untrusted quantum cloud environments.

Overall, existing defenses address different aspects of the compiler trust problem, including circuit reconstruction, intellectual property protection, encoding leakage, and hardware tampering. However, most approaches operate independently and focus on specific threat models. This highlights the need for unified compiler-security frameworks capable of coordinating protection mechanisms across the entire quantum compilation pipeline while balancing security, reliability, and computational overhead.

\subsection{Crosstalk: characterization and mitigation}\label{sec:crosstalk}
Among the various sources of error in quantum computing, crosstalk is one of the most significant multi-qubit-level phenomena affecting computational fidelity. Crosstalk arises from unintended interactions between qubits due to spatial proximity, shared control hardware, or simultaneous operations. When qubits are operated concurrently or in close succession, residual interactions can cause quantum information to leak between qubits, resulting in spurious correlations, data leakage, and incorrect computation outcomes~\cite{PhysRevLett.131.210802,9193969}. 

A particularly important form of crosstalk is ZZ crosstalk, which originates from coupling between the computational basis states of one qubit and the energy levels of another qubit~\cite{4}. This interaction introduces unwanted phase shifts that distort logical operations and reduce gate accuracy. The Hamiltonian representation of ZZ crosstalk is given by Equation~(\ref{ham-rep}):

\begin{equation}
\label{ham-rep}
H = \sum_{i} \omega_{i} \sigma_{z}^{i} + \sum_{i < j} J_{ij} \sigma_{z}^{i} \sigma_{z}^{j}
\end{equation}

where $\omega_{i}$ represents the energy level of qubit $i$, $J_{ij}$ denotes the coupling strength between qubits $i$ and $j$, and $\sigma_{z}^{i}$ and $\sigma_{z}^{j}$ are the Pauli-Z operators associated with the respective qubits. The interaction term captures the unwanted ZZ coupling responsible for crosstalk effects.

\subsubsection{Crosstalk characterization}

Accurate characterization of crosstalk is essential for understanding and mitigating correlated quantum errors. Crosstalk error rates are commonly extracted using two techniques~\cite{5}: Idle tomography (IDT) and simultaneous randomized benchmarking (SRB). IDT analyzes circuits containing idle gates, where selected qubits remain inactive during execution, allowing interference effects to be isolated and studied~\cite{2019APS..MARP35006B}. Using both no-drive and drive circuits, IDT enables the extraction of Hamiltonian, stochastic, and affine error rates~\cite{19}. Hamiltonian errors correspond to over- or under-rotations in the bloch sphere representation, stochastic errors represent the probability of Pauli-type gate errors, and affine errors capture additional deviations beyond these standard error models.

In contrast, SRB evaluates gate performance when multiple operations are executed concurrently~\cite{5}. By comparing gate fidelities under isolated and parallel execution, SRB quantifies the extent to which simultaneous gate operations introduce interference. For example, the error rate of a CNOT gate may increase when another CNOT gate is executed concurrently on a different qubit pair, revealing the presence of crosstalk or correlated noise effects.

\subsubsection{Impact across quantum technologies}

Several studies have demonstrated the detrimental impact of crosstalk on quantum circuit fidelity~\cite{18,20,812634,PRXQuantum.3.020301}. Feng \textit{et al.}~\cite{64} identify photon scattering during entanglement generation as one important source of crosstalk that can degrade stored quantum information. The severity of crosstalk depends on factors such as qubit count, processor topology, and qubit connectivity, with highly connected systems exhibiting more complex interference patterns and greater challenges in error isolation.

The manifestation of crosstalk differs significantly across hardware technologies. In superconducting qubit systems, crosstalk primarily arises from microwave pulse interference, capacitive or inductive coupling between nearby qubits, and overlapping control wiring~\cite{PRXQuantum.3.020301}. Dense chip layouts and parallel gate operations further amplify correlated noise, particularly during multi-qubit operations. To mitigate these effects, software-level schedulers integrated into compilers such as IBM Qiskit and Rigetti Quilc can serialize conflicting operations, while hardware-level solutions including tunable couplers, improved shielding, and optimized chip layouts help suppress unwanted interactions. Quantum error correction codes, particularly surface codes, further assist in mitigating residual correlated noise.

In ion-trap quantum systems, crosstalk mainly results from stray laser light, imperfect beam focusing, and unintended excitation of motional modes~\cite{18,20}. Although ion traps benefit from all-to-all qubit connectivity, scaling larger ion arrays increases the likelihood of unintended coupling during simultaneous operations. Mitigation strategies therefore focus heavily on hardware and pulse-engineering techniques such as improved laser focusing, optical shielding, modular trap architectures, and refocusing pulse sequences.

Crosstalk becomes particularly problematic in multi-programming or multi-tenant quantum environments where several users share the same quantum processor. Ash-Saki \textit{et al.}~\cite{19} demonstrate using benchmark algorithms such as Grover's search and quantum Fourier transform that adversarial CNOT operations executed concurrently with victim circuits can reduce the probability of correct outputs by up to 15.32\%. Their analysis uses TVD to quantify deviations between noisy and ideal circuit outcomes. Overall, crosstalk remains a major obstacle to reliable and fault-tolerant quantum computation, highlighting the need for platform-specific mitigation techniques combining hardware engineering, intelligent scheduling, and accurate device characterization.

\begin{figure*}[htbp]
    \centering
        \includegraphics[scale=0.17]{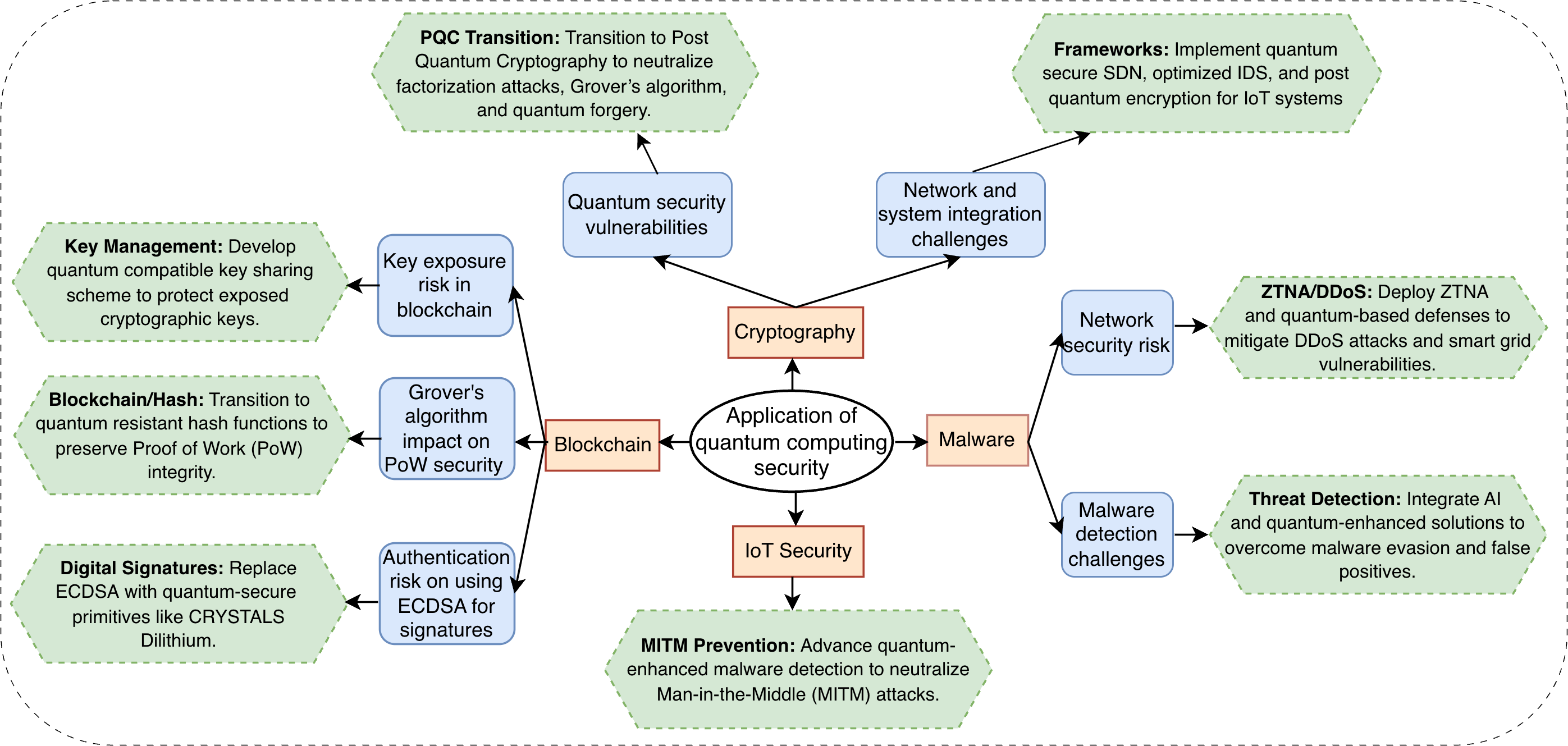}
       \caption{Mapping of quantum-induced cybersecurity issues and their solutions across key domains.}
    \label{fig:error-mitigation}
\end{figure*}

\subsubsection{Mitigation strategies}

Crosstalk mitigation has become an important research direction because unintended qubit interactions directly affect the accuracy, scalability, and reliability of quantum systems~\cite{marquer2025qaiccc,jirovec2025exchange,kang2025crosstalk,zhong2025cycle,liu2025performance,kang2025time,taravati2025space}. Existing mitigation approaches span both hardware and software layers, ranging from intelligent qubit allocation to architectural redesign and noise-aware compilation.

One widely explored direction is noise-aware qubit allocation. Harper \textit{et al.}~\cite{1} propose reinforcement (RL) learning based qubit placement algorithms that dynamically learn device noise characteristics and optimize qubit-to-hardware mappings to reduce crosstalk and gate errors. Similarly, spectator qubits are used to passively monitor fluctuations caused by nearby active qubits, allowing crosstalk effects to be detected without directly interfering with the computation~\cite{1,17}. Seo \textit{et al.}~\cite{17} further propose a two-step mitigation framework consisting of quantum pre-processing, which applies local unitary transformations before measurement, and classical post-processing, which corrects measurement outcomes using known noise characteristics.

Several techniques also focus on reducing interference through circuit placement and characterization. Circuit separation physically spaces concurrently executing circuits to minimize unwanted interactions between neighboring qubits~\cite{4}. Idle tomography characterizes how inactive qubits are affected by nearby operations and feeds this information into crosstalk simulators for proactive mitigation~\cite{28}. Additional approaches such as buffer qubits insert unused qubits between active regions to act as protective barriers against interference, while empirical noise characterization and simulation methods model spatial and temporal crosstalk patterns to support data-driven mitigation strategies.

For shared quantum systems, Marquer \textit{et al.}~\cite{marquer2025qaiccc} propose qubit allocation for inter-circuit crosstalk countermeasure (QAICCC), a scalable and security-aware protocol for multi-tenant quantum environments. QAICCC constructs detailed empirical crosstalk maps covering multiple interaction types and allocates qubits to minimize worst-case inter-circuit interference while maintaining high hardware utilization. Kumar \textit{et al.}~\cite{kumar2025context} introduce a context-switching-based defense mechanism, where quantum programs are executed across multiple co-running contexts to reduce consistent exposure to malicious or noisy neighboring programs. Their multi-programming with frequent context switching strategy further uses statistical techniques such as the Hold-Out method and Hellinger distance to detect abnormal output divergence caused by crosstalk attacks.

Other mitigation strategies are tailored to specific hardware technologies. For superconducting systems, Kang \textit{et al.}~\cite{kang2025crosstalk} propose crosstalk spectating multiple quantum coherences, a single-shot detection technique that uses spectator qubits prepared in modified GHZ states to amplify subtle phase shifts caused by nearby gate operations. Dynamical decoupling is used to suppress unrelated noise sources, enabling accurate crosstalk detection in multi-user environments. Xie \textit{et al.}~\cite{4} address ZZ crosstalk, an always-on interaction common in superconducting qubits, by optimizing pulse shapes, timings, and gate schedules using gradient-based optimization methods.

For trapped-ion systems, Feng \textit{et al.}~\cite{64} propose a dual-type qubit scheme using the $S$ and $F$ hyperfine levels of $^{171}\mathrm{Yb}^{+}$ ions to reduce crosstalk in ion-photon entanglement networks. At the architectural level, Kosen \textit{et al.}~\cite{63} investigate flip-chip quantum architectures, where physical separation between quantum and classical control chips significantly reduces signal interference in densely packed systems. In quantum dot arrays, Jirovec \textit{et al.}~\cite{jirovec2025exchange} propose an exchange crosstalk mitigation framework that characterizes parasitic couplings using avoided crossings and constructs crosstalk matrices to achieve independent control over exchange interactions through virtual barrier voltages.

Beyond hardware-focused techniques, several software-level solutions have also emerged. RC transforms coherent hardware-specific noise into stochastic depolarizing noise by inserting randomly selected Pauli or Clifford gates around two-qubit operations~\cite{62}. This conversion makes errors more uniform and easier to mitigate using existing error correction techniques. Phalak \textit{et al.}~\cite{63} further propose QuantumPUFs, which exploit device-specific noise characteristics such as decoherence and gate error rates to generate unique hardware fingerprints. In addition to providing device authentication and resistance against cloning, QuantumPUFs enable device-aware scheduling and calibration strategies that account for processor-specific crosstalk behavior.

\subsubsection{Synthesis}
Existing crosstalk mitigation techniques reveal an important trade-off between mitigation precision and scalability. Approaches such as RL-based qubit allocation and QAICCC achieve strong performance by relying on detailed empirical characterization of hardware behavior, but the characterization overhead grows rapidly with increasing qubit count. In contrast, techniques such as randomized compiling and context-switching reduce reliance on detailed device models by introducing statistical robustness through randomization and execution diversity. Hardware-focused solutions, including improved architectures, pulse engineering, and spectator-based detection, provide more direct suppression of interference but often require specialized device support. Overall, effective crosstalk mitigation will likely require a combination of hardware-aware scheduling, accurate noise characterization, architectural improvements, and software-level error suppression techniques to support scalable and reliable quantum computation.

\section{Quantum computing for cybersecurity applications}
\label{S4}
Although quantum computing is still evolving and faces several reliability and security challenges, it remains a highly promising computational paradigm with capabilities that are difficult to achieve using classical systems alone. Its potential applications in optimization, simulation, cryptography, and ML have motivated researchers to explore how quantum technologies can also strengthen cybersecurity and address emerging security challenges. Figure~\ref{fig:error-mitigation} maps all security issues induced by quantum computing and their solutions across key domains. The light orange rectangle of the figure represents the key domains affected by quantum computing. The blue rounded rectangle of the figure represents the key issues caused by the inclusion of quantum computing in its respective domains. And finally, the green hexagon represents the solutions to the aforementioned issues.

\begin{table*}[!tb]
\centering
\caption{Comparison of quantum-based security solutions, applications, and trade-offs}
\label{tab:quantum_security_comparison}
\footnotesize
\renewcommand{\arraystretch}{1.0}
\setlength{\tabcolsep}{4pt}
\newcommand{\tildot}{\,\textbullet\;}
\begin{tabularx}{\textwidth}{@{}p{2cm}XX@{}}
\toprule
\textbf{Solution} & \textbf{Applications} & \textbf{Trade-off / limitation} \\
\midrule
PQC
& 
\begin{itemize}
    \item Quantum Cryptography: Resistance against quantum attacks
    \item Blockchain: Replacing ECDSA and SHA-256 with PQ-safe alternatives
    \item Software Security: Long-term software protection (LWE, Merkle Trees, McEliece)
    \item IoT Security: TLS handshakes, key encapsulation mechanisms (KEMs), and DSAs
    \item Cyber Security: K-PAKE (CRYSTALS-Kyber), satellite and smart meter privacy
\end{itemize}
&  
\begin{itemize}
    \item High bandwidth and latency overhead in IoT due to large key sizes
    \item Large storage requirements for hash-based signatures
    \item High computational cost for lattice-based schemes
\end{itemize}
\\
%\hline
QKD
& 
\begin{itemize}
    \item Quantum Cryptography: Secure key exchange (BB84), fiber networks, 5G/6G
    \item IoT Security: Challenge-response mechanisms (Quantum Guard)
\end{itemize}
& 
\begin{itemize}
    \item Distance limitation ($\approx 70$ km) due to fiber loss
    \item Requires specialized and costly hardware
    \item Highly sensitive to environmental conditions
\end{itemize}
  \\
%\hline
QML \& deep learning
& \begin{itemize}
    \item Malware Detection: Hybrid quantum convolutional neural networks (QCNN), recurrent neural network (RNN), quantum support vector machines (QSVM), variational quantum circuit (VQC)
    \item IoT Security: Device-specific malware detection
    \item Cyber Security: Threat detection using quantum kernels
\end{itemize}
& 
\begin{itemize}
    \item Lower accuracy in some benchmarks vs classical models
    \item High resource demand and scalability issues
\end{itemize}
\\
%\hline
Quantum optimization \& search
& \begin{itemize}
    \item Blockchain: Grover’s for mining, annealing for optimization
    \item Malware Detection: Suspicious data processing
    \item Cyber Security: Password recovery and cost optimization
\end{itemize}
& 
\begin{itemize}
    \item Requires multiple quantum processors
    \item Mostly theoretical due to lack of scalable hardware
\end{itemize}
\\
%\hline
Secure protocols \& architectures
& 
\begin{itemize}
    \item Blockchain: quantum ledger technology (QLT), Commit-Delay-Reveal
    \item Malware / Smart Grid: Zero Trust Architecture with quantum reinforcement learning (QRL)
    \item Cyber Security: RoseCliff authentication
\end{itemize}
& 
\begin{itemize}
    \item Operational delays (e.g., fund locking)
    \item Requires major architectural redesign
    \item Interim reliance on classical AES-256
\end{itemize}
\\
\bottomrule
\end{tabularx}
\end{table*}

\subsection{Quantum cryptography and post-quantum security}
\label{sec:crypto}

Cryptography secures information by transforming plaintext into ciphertext, enabling secure communication and storage of sensitive data. Quantum cryptography extends this concept by leveraging quantum mechanical principles such as superposition and entanglement to establish secure communication channels that are resistant to eavesdropping and future quantum-enabled attacks. Unlike classical cryptographic systems, quantum cryptography relies on qubits and quantum state transmission to provide theoretically secure communication.

However, the rapid progress of quantum computing also threatens many widely used classical cryptographic schemes. Quantum algorithms such as Grover's algorithm reduce the effective security strength of symmetric encryption and hash functions by providing quadratic search speedups, effectively halving the security level of schemes such as AES and blockchain-related hash mechanisms~\cite{27,96-blockchain}. More critically, Shor's algorithm enables efficient integer factorization and discrete logarithm computation, threatening public-key systems such as RSA and ECDSA~\cite{29}. The compromise of ECDSA signatures can lead to severe consequences in blockchain systems, including signature forgery and double-spending attacks.

QKD has emerged as one of the most prominent approaches for secure communication in the quantum era. QKD transmits quantum states, typically photonic qubits, to establish shared secret keys between communicating parties. However, practical deployment of QKD faces several limitations. Quantum states are highly sensitive to environmental disturbances such as temperature fluctuations, vibrations, and electromagnetic interference, which can introduce errors and cause loss of quantum information~\cite{48}. In addition, the supporting infrastructure required for scalable quantum cryptography, including photon sources, optical fibers, detectors, and quantum repeaters, is still under active development and remains difficult to integrate efficiently with existing classical systems.

Another major challenge is the limited operational range and performance of current QKD systems. Zbinden \textit{et al.}~\cite{49} report that detector noise and optical fiber losses restrict practical QKD transmission distances to approximately 70~km while also affecting the achievable data rate and quantum bit error rate (QBER). These constraints continue to limit large-scale deployment of quantum cryptographic systems in real-world communication networks.

Overall, while quantum computing introduces serious threats to existing cryptographic infrastructure, it also provides the foundation for new secure communication mechanisms such as QKD and post-quantum cryptography. As quantum technologies continue to mature, the development of scalable, quantum-resistant cryptographic systems will become essential for long-term cybersecurity.

\subsubsection{Solutions}
To address the threat posed by quantum computers to existing cryptographic systems, researchers are developing quantum-resistant security mechanisms that combine PQC with quantum cryptography techniques such as QKD. Institutions including NIST, ENISA, and NATO are actively standardizing and evaluating these approaches to support a gradual transition toward quantum-secure communication systems.

\textbf{Post-quantum cryptography:}
Arslan \textit{et al.}~\cite{27} emphasize the need for cryptographic schemes resilient against quantum-enabled attacks. Current PQC approaches rely on mathematical problems believed to remain difficult even for quantum computers, including lattice-based cryptography, hash-based schemes, code-based cryptography, and multivariate quadratic polynomial systems. These approaches aim to replace vulnerable classical public-key systems while remaining compatible with existing communication infrastructure.

To strengthen digital signatures against quantum attacks, Boney \textit{et al.}~\cite{29} propose the use of Chameleon hash functions and the random oracle model to transform classical signature schemes into quantum-secure variants. Their approach considers quantum queries over superposed messages and signatures, enabling secure authentication even in the presence of quantum adversaries.

\textbf{Quantum key distribution:}
Quantum cryptography primarily operates through QKD, which enables two parties to securely generate and exchange cryptographic keys using quantum states~\cite{8527822}. Because any eavesdropping attempt disturbs the transmitted quantum states, QKD can detect interception attempts during communication.

One of the most widely used QKD protocols is BB84~\cite{9750206}, which detects eavesdropping through measurement disturbances. Bennett \textit{et al.}~\cite{48} further improve the practicality of QKD by proposing an auto-balanced interferometer using Faraday mirrors, eliminating the need for precise polarization alignment while achieving high visibility for fiber-optic deployment.

A major limitation of practical QKD systems is the QBER, which measures the fraction of incorrect bits during transmission:
\[
\text{QBER} = \frac{\text{number of errors}}{\text{total number of bits transmitted}}
\]
High QBER values may indicate poor transmission quality or the presence of an eavesdropper~\cite{49}. Zbinden \textit{et al.}~\cite{49} show that detector noise and fiber losses currently limit practical QKD transmission distances to approximately 70 km.

To improve practical deployment, Elliot \textit{et al.}~\cite{51} developed the defense advanced research projects agency (DARPA) quantum network, the first network to provide end-to-end security using high-speed QKD integrated with IPSec VPN infrastructure. In this architecture, QKD-generated keys periodically replace conventional encryption keys, while AES is used for data encryption. More recently, QKD has also been proposed for integration into emerging 5G~\cite{61} and 6G~\cite{60} communication systems to strengthen the security of sensitive data in applications such as healthcare and IoT.

\noindent\fbox{%
  \parbox{\dimexpr\columnwidth-2\fboxsep-2\fboxrule\relax}{
 \textbf{Note:} Quantum cryptography, particularly QKD, uses principles of quantum mechanics to securely distribute cryptographic keys using quantum states~\cite{27}. In contrast, PQC does not rely on quantum mechanics, but instead uses mathematical problems believed to be resistant to attacks from both classical and quantum computers~\cite{29}.}%
}

\begin{table*}[!tb]
\footnotesize
\centering
\caption{Threat model for security domains impacted by quantum computing (Section~\ref{S4})}
\label{tableSurvey2}
\renewcommand{\arraystretch}{1.0}

\begin{adjustbox}{width=\textwidth}
\begin{tabular}{%
%p{0.7cm}
p{3.2cm}
p{3.8cm}
p{3.8cm}
p{3.6cm}
p{2.6cm}
}
\toprule
\textbf{Threat domain} & \textbf{Threat vector / cause} &
\textbf{Security impact} & \textbf{Preconditions} & \textbf{Section} \\
\midrule

Quantum cryptography &
Limited transmission distance, noise, and infrastructure immaturity in QKD &
Key loss, increased QBER, reduced practicality &
Long-distance quantum channels; current photonic hardware &
\S\ref{sec:crypto} \\

Classical cryptography &
Shor's and Grover's algorithms &
Broken public-key cryptography; reduced symmetric key strength &
Large-scale quantum computing capability &
\S\ref{sec:crypto} \\

Blockchain security &
Quantum attacks on ECDSA and PoW hash functions &
Signature forgery, double spending, mining centralization &
Quantum access; classical blockchain cryptography &
\S\ref{sec:blockchain} \\

Malware detection &
Increasing malware complexity exceeding classical ML capability &
Reduced detection accuracy; evasion of classical defenses &
Polymorphic/metamorphic malware; large datasets &
\S\ref{sec:malware} \\

Quantum ML reliability &
Noise, decoherence, and lack of mature QEC &
Unstable or inaccurate malware detection models &
Execution on NISQ-era quantum hardware &
\S\ref{sec:malware} \\

IT Infrastructure security &
Quantum-enabled cryptanalysis of encryption protocols &
Compromised data confidentiality and integrity &
Use of classical cryptography in IT systems &
\S\ref{sec:itsecurity} \\

Software security &
Probabilistic behavior and debugging difficulty of quantum programs &
Faulty verification; insecure software deployments &
Quantum program development and execution &
\S\ref{sec:itsecurity} \\

Network intrusion detection &
High-dimensional traffic and evolving attack patterns &
Low detection rates; high false positives &
Large-scale network traffic; classical IDS models &
\S\ref{sec:itsecurity} \\

IoT cryptographic security &
Shor- and Grover-based attacks on IoT key exchange and signatures &
Device impersonation; broken authentication &
RSA/ECC/AES-based IoT protocols &
\S\ref{sec:iot} \\

IoT communication protocols &
Store-now-decrypt-later attacks on TLS/MQTT &
Long-term data exposure; MITM attacks &
Recorded encrypted traffic; future quantum decryption &
\S\ref{sec:iot} \\

IoT firmware integrity &
Quantum breaking of signature-based firmware verification &
Malicious firmware injection &
Quantum-capable attackers; classical signing schemes &
\S\ref{sec:iot} \\

IoT key generation &
Weak PRNG-based key generation &
Complete key compromise &
State compromise or brute-force attacks &
\S\ref{sec:iot} \\

Cybersecurity systems &
Inability of classical threat detection to scale with attack complexity &
Delayed or missed threat detection &
Large data volumes; evolving attack codebases &
\S\ref{sec:cybersec} \\

Authentication systems &
Weak password-based authentication under quantum threat &
Credential compromise; impersonation &
Guessable passwords; weak encryption &
\S\ref{sec:cybersec} \\

Satellite and wireless security &
Incompatibility of classical PAKE and encryption protocols &
Eavesdropping; man-in-the-middle attacks &
Open-air transmission; quantum-capable adversary &
\S\ref{sec:cybersec} \\

\bottomrule
\end{tabular}
\end{adjustbox}
\end{table*}

% \subsubsection{Synthesis}
% PQC and QKD address fundamentally different aspects of the quantum cryptographic challenge. QKD provides information-theoretic security for key exchange but is limited by distance ($\sim$70~km) and infrastructure requirements, making it most practical for high-security point-to-point links. PQC algorithms (lattice-based, hash-based, code-based) offer broader deployability without specialized hardware but face trade-offs in key size, signature size, and computational cost. The DARPA quantum network~\cite{51} demonstrated that QKD can be integrated into existing IPSec VPN infrastructure, suggesting a hybrid deployment model: QKD for high-value channels and PQC for general-purpose security. As standardization accelerates (NIST, ENISA, NATO), the primary challenge shifts from algorithm design to migration planning, transitioning legacy systems without disrupting existing security guarantees.

\subsection{Blockchain security}\label{sec:blockchain}

\subsubsection{Challenges}
Blockchain technology, which relies heavily on cryptographic primitives for decentralized trust and transaction validation, is significantly threatened by advances in quantum computing. The large factorization capability of quantum algorithms such as Shor’s algorithm threatens widely used cryptographic schemes including RSA and elliptic curve cryptography (ECC)~\cite{69-blockchain}. As a result, encrypted digital assets and blockchain transactions may become vulnerable to adversaries capable of recovering private keys, forging signatures, manipulating transactions, and compromising the integrity of decentralized ledgers. This risk is particularly severe for blockchain platforms such as Bitcoin and Ethereum, which rely on ECDSA-based digital signatures~\cite{72-blockchain,73-blockchain,74-blockchain}.

Quantum computing also increases the severity of 51\% attacks~\cite{70-blockchain}. In decentralized blockchain systems, an attacker equipped with sufficiently powerful quantum resources could compromise the private keys of multiple nodes and potentially gain control over the network consensus mechanism. Beyond public-key cryptography, blockchain security also depends heavily on cryptographic hash functions used in proof-of-work (PoW) mechanisms and Merkle tree construction. Adam \textit{et al.}~\cite{71-blockchain} highlight that Grover’s algorithm reduces the complexity of brute-force hash searches from linear to square-root complexity, weakening the security guarantees provided by PoW systems and hash-based verification structures.

In addition to cryptographic vulnerabilities, the emergence of quantum technologies introduces broader security concerns, including the development of malware and attack strategies designed specifically for the quantum era. These evolving threats indicate that blockchain security must adapt not only to quantum cryptanalysis, but also to new categories of quantum-enabled cyberattacks.

\subsubsection{Solutions}
To protect blockchain systems against quantum-enabled attacks, researchers have proposed solutions spanning quantum-safe cryptography, secure migration protocols, blockchain architecture redesign, and quantum-enhanced optimization techniques.

\textbf{Post-quantum cryptographic protection:}
Several studies propose replacing vulnerable cryptographic primitives such as RSA, ECC, and ECDSA with post-quantum cryptographic schemes. Saad \textit{et al.}~\cite{69-blockchain} discuss lattice-based, hash-based, and code-based cryptography as quantum-resistant alternatives for digital signatures, key management, and transaction authentication in blockchain systems. Saeed \textit{et al.}~\cite{70-blockchain} further propose multivariate polynomial-based cryptographic techniques as part of PQC integration for blockchain security.

Adam \textit{et al.}~\cite{71-blockchain} extend this direction by recommending the replacement of SHA-256 and SHA-512 with SHA-3 variants to improve resistance against Grover’s algorithm. They note that the effective security strength of SHA-256 is significantly reduced under quantum attacks, whereas SHA-512 retains stronger post-quantum security guarantees. Their work also includes simulations using IBM Qiskit and Google Cirq to estimate the practical impact of quantum attacks on blockchain cryptography.

\textbf{Quantum-safe blockchain protocols and architectures:}
To securely transition existing blockchain systems toward quantum-resistant infrastructures, Stewart \textit{et al.}~\cite{68-blockchain} propose a commit-delay-reveal protocol for migrating vulnerable ECDSA keys to quantum-safe keys. The protocol operates in three phases: commit, delay, and reveal. Initially, the user commits a hash linking the existing public key with a quantum-resistant public key without revealing either key. During the delay phase, funds remain locked for a security period to prevent quantum attackers from exploiting exposed keys. Finally, the reveal phase transfers control to the quantum-safe key while verifying the integrity of the earlier commitment. This approach helps reduce the risk of fund theft during migration to post-quantum blockchain systems.

At the infrastructure level, Raheman \textit{et al.}~\cite{51} introduce quantum-safe ledger technology (QLT), a framework designed to protect distributed ledger technologies and cryptocurrency exchanges against both classical and quantum cyberattacks. QLT relies on zero vulnerability computing, which eliminates third-party permissions and isolates blockchain nodes to improve security against reverse engineering and unauthorized access.

\textbf{Quantum-enhanced blockchain operations: }
Beyond defensive mechanisms, quantum computing is also being explored to improve blockchain efficiency and optimization. Carrascal \textit{et al.}~\cite{52} apply the VQE combined with differential evolution (DE) to solve optimization problems associated with cryptocurrency systems. DE evolves candidate solutions using mutation, crossover, and selection operations to improve convergence toward global minima.

Quantum algorithms are also being explored for accelerating blockchain analytics and financial optimization. Zhou \textit{et al.}~\cite{72-blockchain} propose the use of quantum monte carlo and quantum amplitude estimation, which reduce the number of samples required for estimation tasks and provide quadratic speedup compared to classical monte carlo approaches. This improvement significantly accelerates probabilistic estimation and blockchain-related analytical workloads.

Similarly, Naik \textit{et al.}~\cite{73-blockchain} formulate portfolio optimization as a quadratic unconstrained binary optimization (QUBO) problem  as in Equation~(\ref{QUBO}) suitable for quantum annealers such as D-Wave systems:
\begin{equation}
    \label{QUBO}
    \min_{x \in \{0,1\}^n} \left( x^\top Q x - \lambda x^\top r \right)
\end{equation}
where the formulation balances portfolio risk and expected return. The QUBO model can further be transformed into an Ising Hamiltonian representation for execution on quantum annealing hardware.

\textbf{Future quantum-native blockchain systems:}
Tessler \textit{et al.}~\cite{74-blockchain} discuss the long-term possibility of quantum-native cryptocurrencies such as Qubitcoin, which leverage concepts including entanglement, the no-cloning theorem, and quantum error correction. Such systems aim to prevent double-spending attacks and potentially remove the need for conventional Proof-of-Work mining. The authors also emphasize the importance of adopting post-quantum mining strategies and hybrid mining schemes to prevent quantum-enabled miners from dominating blockchain networks.

\subsection{Malware detection}\label{sec:malware}
\subsubsection{Challenges}
Malware is evolving faster than traditional detection systems can adapt, especially due to increasing use of obfuscation techniques and complex code structures~\cite{23}. Quantum-enhanced deep learning has been proposed to address this issue, as quantum models can process large and complex feature spaces more efficiently. Wiene \textit{et al.}~\cite{wiebe2015quantumdeeplearning} discuss quantum-assisted training of deep neural networks using deep restricted Boltzmann machines (dRBMs) and Full Boltzmann Machines, which could support advanced malware analysis. However, integrating quantum computing with ML introduces significant practical challenges, including quantum noise, decoherence, and the immaturity of quantum error correction mechanisms~\cite{56,57,23}.

Researchers have also highlighted emerging malware-related threats such as advanced persistent threats (APTs), ransomware, phishing, polymorphic malware, and metamorphic malware~\cite{56,57,75-mal_det,76-mal_det}. Fake usernames generated by malicious actors~\cite{59} and DGA-based botnets designed to evade blacklist-based detection~\cite{66} further complicate cybersecurity defense. Smart grids introduce additional risks because malware-infected devices can disrupt critical infrastructure~\cite{78-mal_det}. Hawawreh \textit{et al.}~\cite{77-mal_det} further note that zero trust architecture is still not effectively deployed in operational technology (OT) environments of smart grids.

\subsubsection{Solutions}
Several quantum and hybrid quantum-classical approaches have been proposed to improve malware detection accuracy, scalability, and real-time responsiveness.

Ciaramella \textit{et al.}~\cite{23} introduced a Hybrid-QCNN framework for Android malware detection by combining quantum convolutional layers with classical convolutional neural network (CNN) layers. The quantum convolutional layer performs feature extraction in reduced-dimensional quantum feature spaces, enabling efficient malware classification while using significantly smaller input representations compared to traditional CNN architectures such as VGG16 a popular 16-layer CNN architecture designed for image classification:
\begin{flushleft}
\begin{minipage}{0.45\textwidth}
\begin{align*}
\text{Hybrid-QCNN} &= \text{Quantum Convolutional Layer} \\
&\quad + \text{Classical CNN Layers}
\end{align*}
\end{minipage}
\end{flushleft}

This study demonstrates that quantum-assisted feature extraction can reduce computational overhead while preserving high detection accuracy.

Beyond CNN-based models, Mittal \textit{et al.}~\cite{58} proposed RNN based malware detection models capable of learning temporal dependencies in malware behavior. By analyzing sequential data such as API calls and network traffic flows, the proposed system can identify evolving malware patterns that are difficult to capture using static analysis techniques. The authors reported high detection rates for zero-day malware attacks.

Several works also combine quantum machine learning with conventional ML pipelines. Azeez \textit{et al.}~\cite{57} integrated QSVM with classical ML models such as random forests and gradient boosting, along with deep learning architectures including CNNs and LSTMs. In the proposed hybrid framework, quantum processors handle quantum data encoding and pattern recognition tasks, while classical processors perform decision-making and real-time threat classification. Similarly, Payares \textit{et al.}~\cite{25} explored QSVM, hybrid QNN, and ensemble approaches for DDoS attack detection in cybersecurity systems.

Bikku \textit{et al.}~\cite{56} proposed a QNN-based malware detection framework that uses quantum Fourier transform (QFT) for feature extraction. QFT transforms classical feature vectors into quantum states, enabling the system to exploit quantum superposition and interference during feature analysis. The extracted features are then processed using a VQC, which acts as the classifier. The circuit parameters are optimized by minimizing the loss function ($\mathcal{L}(\theta)$) as in Equation~(\ref{loss-func}):
\begin{equation}
    \label{loss-func}
    \mathcal{L}(\theta) = \sum_{i=1}^{N} \left( y_i - \hat{y}_i(\theta) \right)^2   
\end{equation}

where \(\theta\) denotes trainable circuit parameters. The framework performs real-time malware classification by distinguishing malicious and benign samples based on learned quantum feature representations. Varalakshmi \textit{et al.}~\cite{79-mal_det} similarly proposed a QNN-based ransomware detection framework that integrates quantum feature extraction, variational quantum classifiers, and continuous feedback mechanisms to improve ransomware identification accuracy.

Quantum learning techniques have also been applied beyond conventional malware binaries. Andreev \textit{et al.}~\cite{59} proposed a quantum machine learning framework for detecting fake usernames generated through chaotic pseudo-random number generators. The approach combines quantum k-means clustering with swap-test-based similarity measurements between quantum states to identify suspicious username patterns. Clustering effectiveness is further evaluated using Silhouette-based analysis to distinguish malicious and benign user groups.

Suryotrisongko \textit{et al.}~\cite{66} extended quantum malware analysis toward domain generation algorithm (DGA) botnet detection using hybrid quantum-classical deep learning architectures. The proposed model embeds pennylane-based quantum layers inside classical Keras architectures, replacing selected classical layers with quantum circuits to improve feature discrimination for botnet detection.

For smart-grid cybersecurity, Hawawreh \textit{et al.}~\cite{77-mal_det} proposed a Zero Trust framework integrating software defined networking (SDN), eigengame-based feature extraction, and QRL. The eigengame module is used to extract uncorrelated features from heterogeneous smart-grid data, improving the robustness of malware detection across different operational zones.

The QRL module models malware detection as a Markov decision process and continuously updates detection policies using the Equation~(\ref{mdp}):
\begin{equation}
    \label{mdp}
    \theta^{P+1} \leftarrow \theta^P + \alpha \nabla \theta^P L(\theta^P)
\end{equation}

where the learning loss ($L$) is defined as:
\begin{equation}
    \label{learning-loss}
    L = \text{MSE}\left(r_t + \gamma \max_a Q(O_{t+1}, a; \theta^T) - Q(O_t, l_t; \theta^P)\right)
\end{equation}

This adaptive framework enables real-time ransomware and malware detection in dynamic smart-grid environments.

Akash \textit{et al.}~\cite{78-mal_det} further proposed a cloud-based malware detection architecture for smart grids that combines QCNN, deep transfer learning, and quantum data encoding techniques. The framework uses pretrained CNN models for initial malware learning and fine-tunes them using device-specific firmware datasets, while QCNN modules improve feature extraction from malware binary images.

\subsection{IT, software, and network intrusion detection}\label{sec:itsecurity}

\subsubsection{Challenges}
The emergence of quantum computing introduces major challenges for information technology, software security, and network protection because many existing security mechanisms rely on cryptographic techniques that can be weakened by quantum algorithms~\cite{nyari2021impact,26}. Suryotrisongko \textit{et al.}~\cite{26} point out that encryption and cryptographic systems must evolve toward quantum-resistant standards, while the inherent instability of qubits and decoherence create additional challenges in implementing reliable quantum systems. The authors also emphasize that quantum communication protocols such as QKD require protection against implementation flaws and protocol-level attacks. Similarly, Akbar \textit{et al.}~\cite{60} discuss that although quantum signal processing can improve throughput and reduce latency in future 6G communication systems, it also introduces new risks related to signal tampering and interception during transmission.

Quantum computing also affects software engineering and software verification practices. Alyami \textit{et al.}~\cite{31} explain that quantum algorithms differ fundamentally from classical algorithms, requiring software engineers to adopt new programming models, frameworks, and verification methodologies. Due to the probabilistic nature of quantum systems, debugging and testing quantum software become significantly more difficult than in classical environments~\cite{31,33,10.1145/3402127.3402131,10.1145/3715885.3715895}. Issel \textit{et al.}~\cite{36} further highlight the challenges of integrating quantum security into SDNs. While SDNs provide flexible and programmable network management, their dependence on classical cryptographic techniques makes them vulnerable to quantum attacks. The authors also note that current quantum hardware lacks sufficient qubits to efficiently solve large satisfiability (SAT) problems required in software verification and symbolic model checking.

Public-key infrastructures (PKIs), which secure web communication and authentication, are also highly vulnerable in the quantum era. Awan \textit{et al.}~\cite{35} point out that quantum computers could break asymmetric cryptographic schemes used in PKIs through efficient factorization and discrete logarithm attacks, threatening secure communication over the internet. As organizations increasingly depend on distributed systems, SDNs, and cloud-based services, protecting data security in programmable and network-centric infrastructures becomes increasingly critical.

In the context of network intrusion detection, existing IDS techniques struggle to cope with the growing complexity and scale of modern network traffic~\cite{81-intrusion,82-intrusion,84-intrusion}. Shen \textit{et al.}~\cite{81-intrusion} observe that traditional ML and meta-heuristic approaches often require large computational resources and become trapped in local optima, limiting their suitability for lightweight and real-time intrusion detection. Soliman \textit{et al.}~\cite{84-intrusion} categorize the major limitations of existing IDS systems into several areas, including dependence on known attack signatures, high false positive rates in anomaly detection, scalability issues under high-volume traffic, computational overhead of AI/ML techniques, and difficulty handling high-dimensional datasets. Traditional IDS systems also struggle to balance detection accuracy with false alarm reduction, particularly in dynamic and evolving attack environments. These limitations indicate the need for more adaptive and computationally efficient intrusion detection approaches capable of handling future quantum-era cyber threats.

\subsubsection{Solutions}

To address the security risks introduced by quantum computing, researchers have proposed solutions spanning post-quantum cryptography, quantum-enhanced intrusion detection, and software verification techniques.

A major focus is the development of quantum-resistant cryptographic algorithms. Alyami \textit{et al.}~\cite{32} and Nosouhi \textit{et al.}~\cite{37} discuss several post-quantum cryptographic approaches designed to replace vulnerable classical public-key systems. Among them, lattice-based cryptography is considered one of the most promising because of its resistance to known quantum attacks. Its security relies on hard mathematical problems such as learning with errors (LWE). Lattice-based methods support encryption, digital signatures, and fully homomorphic encryption, making them suitable for practical post-quantum systems.
% , represented as in Equation~(\ref{lwe}):
% \begin{equation}
%     \label{lwe}
%     \mathbf{b} \equiv \mathbf{A} \cdot \mathbf{s} + \mathbf{e} (\text{mod } q)
% \end{equation}

% where recovering the secret vector \(\mathbf{s}\) from noisy equations remains computationally difficult even for quantum computers. 

Hash-based cryptography is another quantum-resistant approach that constructs digital signatures using hash trees such as the Merkle signature scheme (MSS)~\cite{32,cryptoeprint:2005/192}. The hash-tree construction can be expressed as in Equation~(\ref{hash-tree}):
\begin{equation}
    \label{hash-tree}
    H = H(H(\cdots H(K_1) \parallel H(K_2) \cdots) \parallel \cdots)
\end{equation}

where $H$ is a hash function and a set of \( 2^n \) one-time keys \( (K_1, K_2, \ldots, K_{2^n}) \). These schemes provide strong security against quantum attacks because their protection depends on hash-function resistance rather than factorization problems. However, large signature sizes and one-time key usage remain practical limitations~\cite{37}. In addition, code-based cryptography, such as the McEliece cryptosystem~\cite{32,repka2014overview}, secures communication using error-correcting codes as in Equation~(\ref{sc-ec}):
\begin{equation}
    \label{sc-ec}
    \mathbf{c} = \mathbf{m} \cdot \mathbf{S} \cdot \mathbf{G} \cdot \mathbf{P} + \mathbf{e} 
\end{equation}

where \(\mathbf{S}\), \(\mathbf{G}\), and \(\mathbf{P}\) represent scrambling, generator, and permutation matrices, respectively. These approaches collectively form the foundation of post-quantum cryptography for future IT infrastructures.

Researchers have also explored methods for evaluating and strengthening software security in the quantum era. Authors in~\cite{31,34,35} propose combining the fuzzy analytic hierarchy process and fuzzy technique for order preference by similarity to ideal solution (FTOPSIS) for quantum-aware software security assessment. These methods use fuzzy logic to handle uncertainty in security decision-making and help prioritize the most suitable protection mechanisms for quantum-era software systems.

For software reliability and verification, Ali \textit{et al.}~\cite{33} introduce quantum software engineering, which uses projection-based assertions to compare expected and actual quantum states during program execution. Similarly, Issel \textit{et al.}~\cite{36} propose formal verification techniques that transform software verification into satisfiability (SAT) problems as in Equation~(\ref{sat}):
\begin{equation}
    \label{sat}
    f(x_1, x_2, \dots, x_n) = \bigwedge_{i=1}^{n} \left( x_i \lor \neg x_j \right)
\end{equation}

The SAT formulation is then converted into an optimization problem as in Equation~(\ref{sat-opt}):
\begin{equation}
    \label{sat-opt}
    \min_x \left( x^T Q x + c \right)
\end{equation}

allowing quantum optimization methods to assist software verification. Although current quantum hardware limitations restrict practical deployment, these methods demonstrate how quantum computing may support future software validation and debugging processes.

In the area of network intrusion detection, several quantum-inspired and hybrid quantum-classical IDS models have been proposed. Shen \textit{et al.}~\cite{81-intrusion} introduce the Global-best-guided Quantum-inspired tabu search (GQTS) algorithm, which applies quantum-inspired superposition concepts to improve rule generation and avoid local optima during optimization. The fitness of generated detection rules is evaluated using the Equation~(\ref{fitness-ev}):
\begin{equation}
    \label{fitness-ev}
    \text{fitness} = w \cdot \frac{\text{predict}_{attack}}{N_{attack}} + (1-w)\cdot \frac{\text{predict}_{normal}}{N_{normal}} 
\end{equation}

where \(w\) controls the balance between attack and normal traffic classification. Abreu \textit{et al.}~\cite{82-intrusion} further propose QML-IDS, which integrates VQC, QSVM, and QCNN with classical processing for attack detection. Soliman \textit{et al.}~\cite{84-intrusion} also develop the QVICA-with-EDA framework to improve intrusion detection accuracy and reduce false alarms in high-dimensional network environments. In contrast, Salvakkam \textit{et al.}~\cite{83-intrusion} present the ensemble intrusion detection model for cloud computing using deep learning (EICDL), which combines Text-CNN-based feature extraction with deep neural classification for cloud-based IDS systems.

Overall, current solutions demonstrate that protecting IT systems in the quantum era requires both migration toward post-quantum cryptography and the integration of quantum-inspired intelligence into software verification and intrusion detection. While practical deployment challenges remain, these approaches provide important foundations for building secure and resilient IT infrastructures against future quantum-enabled attacks.

\subsection{IoT security}\label{sec:iot}
\subsubsection{Challenges}
The security of IoT systems faces significant challenges in the quantum era, particularly because many existing IoT security mechanisms rely on cryptographic algorithms that can be weakened or broken by quantum computers. Shor’s algorithm threatens widely used asymmetric cryptographic schemes such as RSA, ECC, and Diffie–Hellman key exchange (DHKE), which are commonly used for key exchange, authentication, and digital signatures in IoT systems~\cite{42,92-iot}. This creates serious risks for applications such as healthcare IoT, smart homes, industrial IoT, and smart cities, where attackers may derive private keys from public information and compromise device authentication. In addition, Grover’s algorithm reduces the effective security strength of symmetric encryption and hash functions by providing quadratic speedup for brute-force search~\cite{92-iot}. For example, AES-128 security is effectively reduced to the level of AES-64, requiring stronger cryptographic configurations such as AES-256 to maintain security against quantum attacks.

Beyond cryptographic weaknesses, protocol-level and infrastructure-level vulnerabilities further increase IoT security risks. Schöffel \textit{et al.}~\cite{38} identify increased TLS handshake latency caused by the large key sizes and ciphertext overhead associated with post-quantum cryptographic algorithms. The authors also highlight the \enquote{Store Now, Decrypt Later} threat model, where attackers capture encrypted IoT communication today and decrypt it in the future once large-scale quantum computers become available. Another major concern is that breaking digital signatures in X.509 certificates can enable attackers to forge Root CA certificates, facilitating man-in-the-middle attacks against IoT servers and communication channels~\cite{38}. Xiong \textit{et al.}~\cite{93-iot} further emphasize that quantum attacks against firmware-signing mechanisms may allow adversaries to inject malicious updates into smart meters and industrial controllers, potentially causing critical infrastructure failures.

Resource limitations and architectural diversity also make IoT systems difficult to secure against quantum threats. El-Latif \textit{et al.}~\cite{45} point out that PRNG-based key generation mechanisms used in industrial IoT systems are vulnerable to state compromise and brute-force attacks, which become more feasible with quantum-assisted computation. In addition, many IoT devices are battery-powered and resource-constrained, making it difficult to deploy computationally intensive PQC algorithms efficiently~\cite{42}. Chawla \textit{et al.}~\cite{39} categorize IoT vulnerabilities across sensing, network, and application layers, including node capture, false data injection, denial-of-service attacks, data tampering, weak authentication, and privacy leakage. These vulnerabilities also extend to blockchain-enabled IoT systems, where quantum attackers may exploit weaknesses in digital signatures to falsify transactions or steal assets within smart-city infrastructures~\cite{94-iot}. Furthermore, polymorphic and metamorphic malware continue to undermine classical detection mechanisms in IoT environments, increasing the need for more adaptive and quantum-resistant security solutions~\cite{80-iot}.

\subsubsection{Solutions}
Securing IoT systems against quantum threats requires a combination of post-quantum cryptography, lightweight encryption mechanisms, and distributed security architectures. Existing research focuses both on adapting classical cryptographic techniques to resist quantum attacks and on integrating quantum-enabled security mechanisms into IoT infrastructures.

Schöffel \textit{et al.}~\cite{38} investigate the use of post-quantum key encapsulation mechanisms and digital signature algorithms (DSAs) for securing TLS communication between IoT devices and servers. Their work mainly focuses on the impact of PQC algorithms on TLS handshake latency, bandwidth usage, and energy consumption, which are critical concerns for resource-constrained IoT devices. The bandwidth overhead associated with PQC-enabled TLS communication is expressed as in Equation~(\ref{bw-pqc-tls}):
\begin{equation}
    \label{bw-pqc-tls}
    bw_{\text{DSA}} = \lvert pk_C \rvert + \lvert sig_C \rvert + \lvert pk_S \rvert + \lvert sig_S \rvert + \lvert pk_{CA} \rvert + 3 \cdot \lvert sig_{CA} \rvert
\end{equation}

where \(pk\) and \(sig\) denote the public-key and signature sizes of the client, server, and certificate authority. Their study highlights the practical trade-off between quantum-resistant security and communication overhead in IoT environments.

Several researchers~\cite{39,40,41,42,43,44,45} further investigate lightweight and quantum-resistant cryptographic approaches for IoT systems, including lattice-based, hash-based, and code-based encryption and digital signature schemes. These studies emphasize that future IoT systems will require security mechanisms that not only resist quantum attacks but also remain computationally feasible for low-power and memory-constrained devices.

Volya \textit{et al.}~\cite{46} propose a quantum walk-based encryption framework for IoT environments. Their approach divides the original data into smaller blocks and applies quantum walk (QW)-based permutation and substitution operations to each block individually. QW-derived pseudo-random number generators (PRNGs) and P-Boxes are used to improve confusion and diffusion during encryption. After processing, the encrypted blocks are recombined to reconstruct the final ciphertext. The framework demonstrates how quantum-inspired randomness can strengthen encryption quality for IoT communication and image protection.

To address malware threats in IoT systems, Khan \textit{et al.}~\cite{80-iot} introduce Quantum Guard, a distributed malware detection framework designed for resource-constrained IoT devices. In this architecture, lightweight operations such as memory scanning and feature extraction are performed locally on IoT devices, while computationally intensive quantum operations are offloaded to remote quantum processors. The framework employs QKD-based challenge-response mechanisms for secure communication and utilizes Grover’s algorithm to accelerate malware analysis and pattern matching. This distributed design reduces the processing burden on IoT devices while improving detection capability against advanced malware threats.

Overall, existing IoT security solutions increasingly combine lightweight post-quantum cryptography, quantum-enhanced encryption, and distributed malware defense architectures to address the security challenges introduced by quantum computing. These approaches aim to balance strong security guarantees with the practical resource limitations of IoT environments.

\subsection{Cybersecurity}\label{sec:cybersec}
The transition to a quantum-enabled landscape necessitates a paradigm shift in cybersecurity, as classical cryptographic foundations and authentication protocols face obsolescence. This section categorizes the emerging threats and the multifaceted quantum-resistant solutions proposed in recent literature.

\subsubsection{Challenges}
Umejiaku \textit{et al.}~\cite{85-cyber} discuss the limitations of traditional text-based password authentication systems, where weak and predictable passwords increase the risk of credential compromise and unauthorized access. Authors in~\cite{85-cyber,86-cyber,87-cyber,88-cyber} further highlight that the emergence of quantum computing threatens existing encryption mechanisms because quantum algorithms such as Shor’s algorithm can potentially break widely used public-key cryptographic techniques. This creates serious concerns for the long-term security of current authentication and encryption infrastructures.

Traditional cyber threat detection systems also face increasing challenges in handling the growing complexity and volume of modern cyberattacks. The authors in~\cite{90-cyber,99-cyber,100-cyber,9896970,SURYOTRISONGKO2022223} identify that classical detection frameworks often fail to process large-scale threat data efficiently in real time, resulting in delayed threat identification and slower security responses. Consequently, researchers suggest integrating quantum computing and quantum machine learning techniques to improve cyberattack detection capabilities.

Satellite communication systems introduce another major cybersecurity concern in the quantum era. Since satellite communications occur over open environments, they remain vulnerable to eavesdropping and man-in-the-middle attacks. Yang \textit{et al.}~\cite{89-cyber} explain that conventional cryptographic protocols and existing password authenticated key exchange (PAKE) schemes are not designed for quantum-secure satellite environments and often require multiple rounds of communication, increasing latency and resource consumption. These limitations further emphasize the need for quantum-resilient cybersecurity mechanisms.
.
\subsubsection{Solutions}
To address emerging quantum-era cybersecurity threats, researchers have proposed solutions based on post-quantum cryptography, hybrid authentication systems, quantum-aware threat analysis, and quantum-enhanced cyber threat detection frameworks.

To overcome weaknesses in traditional password-based authentication systems, Umejiaku \textit{et al.}~\cite{85-cyber} proposed the RoseCliff algorithm, a dual-authentication framework designed to strengthen password security using dynamic credentials and modern encryption techniques. The framework consists of components such as input detection, password splitting, one-time numbers (OTN), dynamic ciphertext generation, and authentication. Hybrid encryption is employed by combining AES-256 with Diffie--Hellman key exchange. AES-256 encrypts the dynamically generated password, while Diffie--Hellman uses password segments and OTN values to generate shared random values for secure communication. Although the framework relies on current encryption standards, the authors emphasize the future need for quantum-resistant cryptographic algorithms. Veer \textit{et al.}~\cite{86-cyber} similarly argue that cybersecurity professionals and ethical hackers should incorporate quantum-aware penetration testing techniques, including simulations of attacks based on Shor’s algorithm and AI-driven attack models, to identify vulnerabilities in both classical and post-quantum systems.

Bounceur \textit{et al.}~\cite{87-cyber} proposed another password-related security solution by partitioning Grover’s algorithm across multiple quantum processors combined with classical brute-force computation. In this approach, the initial qubits of the password search space are explored using Grover’s algorithm to leverage quantum speedup, while the remaining search space is processed using parallel classical computation. The partial results from quantum workers are aggregated by a classical controller to determine the final password candidate. This work demonstrates how hybrid classical-quantum computation can be used to address authentication-related security problems.

Several researchers also focus on post-quantum cryptographic frameworks for critical infrastructures. Zhang \textit{et al.}~\cite{88-cyber} proposed a lattice-based privacy-preserving cryptosystem for smart grids using the hardness of short integer solution (SIS) and inhomogeneous SIS (ISIS) problems. The framework encrypts smart-meter power consumption data using lattice-based public-key cryptography to ensure confidentiality, authenticity, and resistance against passive, active, and collusion-based attacks. Similarly, Yang \textit{et al.}~\cite{89-cyber} introduced K-PAKE, a one-round post-quantum Password Authenticated Key Exchange protocol for satellite communications. The framework relies on CRYSTALS-Kyber and module learning with errors (MLWE) to provide resistance against quantum attacks. To further strengthen security, the protocol integrates smooth projective hash functions with gray zones (SPHFwGZ), PBKDF2 with SHA-256, and zero-knowledge proofs to prevent information leakage during authentication.

Azeez \textit{et al.}~\cite{90-cyber} proposed a hybrid cyber threat detection framework integrating quantum computing with artificial intelligence and quantum machine learning. The proposed system combines QSVM and QNN models with classical AI techniques such as random forests, gradient boosting machines, CNNs, and deep learning methods for network traffic analysis and real-time threat detection. Quantum annealing using D-Wave systems is employed for optimization, while amplitude encoding is used to map classical datasets into quantum states for efficient processing. The framework distributes responsibilities between quantum processors, which perform data encoding and pattern recognition, and classical AI systems, which handle decision making and threat classification. Overall, these studies demonstrate that combining post-quantum cryptography, quantum computing, and AI-driven security frameworks can significantly strengthen cybersecurity systems against future quantum-enabled attacks.

\section{Benchmarks used}\label{benchmark}

\begin{table}[!tb]
\centering
    \caption{\textbf{Benchmarks used in studies in focus}}
    \label{tab:benchmark}
\begin{tabular}{ll}
\textbf{Benchmark used} & \textbf{Reference}   \\ \hline
ibmq\_qasm\_simulator                       &    \cite{45,mafi2024quantum,abouelela2020quantum,shenoy2020demonstration}                               \\
ibmq\_manila                                &    \cite{45,12,10629340,shaffar2022experimental}                      \\
ibmq\_ehningen                              &    \cite{37,9951287,9951284}                        \\
ibmq\_hanoi                                 &    \cite{1,6,ikeda2023first,10.1007/978-3-031-40725-3_27}                     \\
ibmq\_guadalupe                             &    \cite{2,10.1007/978-3-031-40725-3_27,das2021adapt,consiglio2024variational}                \\
ibmq\_jakarta                               &    \cite{2,10629340,shi2023error}                   \\
ibmq\_essex                                 &    \cite{5,14,Zaman_Shin_IFM_IBMQ}                               \\
ibmq\_vigo                                  &    \cite{7,10,10.1063/5.0046930}                               \\
ibmq\_casalanca                             &   \cite{7,21}                                \\
ibmq\_melbourne                             &   \cite{7,10,Geller_2021,moradi2022clinical,kuzmak2021preparation}                               \\
ibmq\_tokyo                                 &   \cite{7,jo2022simulating,oancea2025optimizing,das2022foresight,ye2024mutual}                               \\
ibmq\_lagos                                 &   \cite{8,12,24,10629340,garberoglio2025enhanced}                               \\
ibmq\_almadan                               &   \cite{10}                                \\
ibmq\_lima                                  &   \cite{12,22,10629340,sym15010062}                                \\
ibmq\_yorktown                              &   \cite{14,10.1145/3489517.3530400,9645257,9611142}                                \\
ibmq\_nairobi                               &   \cite{18,10629340,javanmard2022quantum,khare2023parallelizing}                                \\
ibmq\_london                                &   \cite{28,slimen2021discrete,coggins2020software}                               \\
ibmq\_burlington                            &   \cite{28,ovaskainen2025quantumsoftwaresecuritychallenges,perelshtein2020large}          \\
ibmq\_perth                                 &   \cite{10629340,10.1063/5.0194993,pandey2022quantum}         \\
ibmq\_quito                                 &   \cite{10629340,li2024qust,singhal2024performance}         \\
\end{tabular}
\end{table}

In this section, we briefly review the various simulation platforms and benchmarks employed in the literature, as summarized in Table~\ref{tab:benchmark}. These benchmarks encompass both quantum simulators and physical quantum computers used to validate experimental results. For instance, \textit{ibmq\_qasm} is utilized in \cite{45} to simulate quantum circuits for a quantum random number generator (QRNG), while the same study employs \textit{ibmq\_manila} to execute these circuits on physical hardware. Unlike the simulated environment, the Manila processor enables the generation of true random numbers by leveraging inherent physical stochasticity. 

Beyond QRNG applications, \textit{ibmq\_manila} is used in \cite{12} to extract control pulse information essential for evaluating power side-channel attacks. Other IBM devices, such as \textit{ibmq\_hanoi}, are employed in \cite{1,6} to validate diverse quantum computing techniques. Similarly, \cite{2} utilizes \textit{ibmq\_guadalupe} and \textit{ibmq\_jakarta} to assess error mitigation strategies in real-world environments, with the latter also serving as the primary evaluation platform in \cite{11}.

The specific architectural constraints of certain devices often dictate their use in research. The \textit{ibmq\_essex} backend, characterized by its unique qubit layout and connectivity, is used in \cite{5} to analyze and mitigate crosstalk. In \cite{14}, \textit{ibmq\_essex} is further utilized to implement the Bernstein-Vazirani algorithm to test hybrid quantum-classical error mitigation schemes.

Several studies focus on the temporal stability and connectivity of quantum systems. In \cite{7}, devices including \textit{ibmq\_vigo}, \textit{ibmq\_casablanca}, \textit{ibmq\_melbourne}, and \textit{ibmq\_tokyo} are used to analyze the temporal variation of error metrics such as $T_1$ relaxation time and $U_3$ gate errors. The analysis of ibmq\_vigo specifically reveals that error rates fluctuate over time and across different qubits, impacting the overall security and resilience of the system. Furthermore, \cite{10} uses \textit{ibmq\_vigo} to illustrate coupling constraints during the mapping of logical to physical qubits, while \textit{ibmq\_casablanca} is used in \cite{21} to demonstrate how specific connectivity patterns influence gate application. To address security concerns, \cite{10} also leverages \textit{ibmq\_melbourne} to demonstrate a split compilation method designed to protect intellectual property from theft.

Finally, a variety of other IBM backends support specialized security and error analyses. Studies \cite{8,12,24} employ \textit{ibmq\_lagos} and \textit{ibmq\_lima} to demonstrate the efficacy of their proposed countermeasures. Specifically, \textit{ibmq\_lima} is used in \cite{22} to evaluate crosstalk impacts on Grover’s algorithm under adversarial scenarios. Additionally, \textit{ibmq\_almaden} is utilized for circuit execution in \cite{10}, while \textit{ibmq\_yorktown} is employed in \cite{14} for quantum amplitude estimation. Lastly, \cite{18} applies the simultaneous randomized benchmarking protocol on \textit{ibmq\_nairobi} to characterize and analyze inter-qubit crosstalk.

%Table~\ref{Table1} summarizes the IBM Quantum simulators and real quantum hardware used across the surveyed studies. These platforms range from noise-free simulators (ibmq\_qasm\_simulator) to physical processors of varying qubit counts and topologies (e.g., 5-qubit Manila, 27-qubit Hanoi, 16-qubit Guadalupe). Several patterns emerge from this landscape. First, the most frequently used backends, Manila, Lagos, Lima, and Nairobi, are small-scale (5--7 qubit) devices, reflecting the reality that most security-relevant experiments require only modest qubit counts but high gate fidelity. Second, studies addressing noise characterization and crosstalk~\cite{7,18,28} tend to use multiple backends (Vigo, Melbourne, Tokyo) to demonstrate that their findings generalize across topologies. Third, the ibmq\_qasm\_simulator serves as a baseline for validating techniques before deployment on noisy hardware, enabling controlled ablation of noise effects. A notable gap is the absence of non-IBM platforms (e.g., Google Sycamore, IonQ, Rigetti) from the surveyed literature, limiting the generalizability of reported results to superconducting architectures.

\section{Open problems and future directions}\label{sec:future}

Our survey reveals several critical gaps that merit focused research effort. We organize these by the two pillars of this study.

\textbf{Security of quantum systems:}
\begin{enumerate}
    \item \textit{Unified compiler security model:} Current defenses against untrusted compilers remain fragmented, as existing approaches such as split compilation, circuit obfuscation, adaptive shot distribution, and TetrisLock each address only specific categories of threats without coordinated integration. A unified framework that classifies compiler threats according to their attack surfaces and systematically maps them to layered countermeasures would enable more principled and effective defense selection.

    \item \textit{Scalable crosstalk characterization:} Techniques such as QAICCC and RL-based qubit allocation require expensive empirical noise characterization that grows with qubit count. As devices scale beyond 100 qubits, efficient characterization methods, perhaps leveraging ML on sparse measurement data, become essential.

    \item \textit{Training challenge in ML-based error mitigation:} ML-based error mitigation techniques, such as ANN-QEM and DAEM, depend on clean and accurate training data. However, obtaining such data is difficult because current quantum systems are inherently noisy. This creates a major challenge, since noisy systems are being used to generate the very data needed to learn how to remove noise. Self-supervised and physics-informed learning approaches, which reduce the dependence on perfectly labeled training data, represent promising but still largely unexplored research directions.

    \item \textit{Cross-platform security evaluation:} Most existing experimental studies discussed in this survey are conducted using IBM Quantum hardware. As quantum computing platforms differ significantly in their architectures and physical implementations, it is still unclear whether the reported vulnerabilities and defense mechanisms are applicable across other technologies such as trapped-ion systems, photonic quantum computers, and neutral-atom platforms. Systematic security evaluation across diverse quantum hardware platforms is, therefore, necessary to understand the generalizability and robustness of existing security solutions.
\end{enumerate}

\textbf{Quantum computing for cybersecurity.}
\begin{enumerate}
    \item \textit{Practical quantum advantage for security applications:} Although hybrid quantum-classical approaches such as QCNN, QSVM, and VQC have demonstrated promising results for cybersecurity tasks, several comparative studies~\cite{75-mal_det} report that these methods may still perform worse than well-optimized classical approaches in certain scenarios. In some cases, the accuracy gap can be around 5\%. This raises an important open question regarding whether current quantum approaches provide a meaningful practical advantage. More rigorous benchmarking studies with fair computational comparisons are required to identify the specific conditions under which quantum methods can outperform classical techniques.

    \item \textit{PQC migration planning for legacy systems:} The threat of \enquote{store-now-decrypt-later} attacks has increased the urgency of transitioning existing systems toward post-quantum cryptography. However, most current research focuses mainly on proposing new algorithms rather than providing practical migration strategies for real-world systems such as PKI infrastructures, TLS protocols, and blockchain networks. Developing hybrid transition frameworks that maintain backward compatibility with existing classical systems remains an important research challenge with significant practical relevance.

    \item \textit{IoT-specific PQC optimization:} Resource-constrained IoT devices face significant challenges when deploying post-quantum cryptographic algorithms because PQC schemes often require larger keys, longer ciphertexts, and higher computational overhead. As a result, there is a need for lightweight PQC implementations specifically optimized for different IoT environments, including medical, industrial, and consumer applications. Balancing security, energy consumption, communication overhead, and device limitations remains an open problem.

    \item \textit{Quantum-safe blockchain coordination:} Transitioning blockchain systems toward quantum-resistant security mechanisms requires replacing widely used cryptographic primitives such as ECDSA and SHA-256 across decentralized networks. However, this is not only a technical challenge but also a coordination and governance problem, since large blockchain ecosystems involve many independent participants and stakeholders. Research on phased migration strategies, incentive mechanisms, and decentralized coordination methods for adopting post-quantum cryptography in blockchain systems is still limited.

    \item \textit{Quantum IDS at scale:} Existing quantum-enhanced intrusion detection approaches, including GQTS, QML-IDS, and QVICA, have primarily been evaluated using relatively small datasets and controlled experimental settings. Their effectiveness in large-scale real-world environments with high-volume network traffic remains uncertain. Scaling these approaches while maintaining detection accuracy, low latency, and meaningful quantum-processing benefits is still an open research challenge.
\end{enumerate}

\section{Conclusion}\label{conclusion}

This survey examined the intersection of quantum computing and cybersecurity from two complementary perspectives: the security of quantum systems themselves and the application of quantum computing to cybersecurity domains. The survey showed that quantum systems introduce several security challenges, including noise exploitation, information leakage, untrusted compilation, and crosstalk, each requiring mitigation strategies at different abstraction levels. Existing defenses span statistical error mitigation techniques, compiler-level transformations, pulse-level optimization, circuit obfuscation, split compilation, and homomorphic encryption. However, no single approach is sufficient to address all security concerns, and practical quantum systems will likely require the integration of multiple complementary techniques to ensure reliability, confidentiality, and secure execution.

At the same time, quantum computing is expected to significantly impact classical cybersecurity infrastructures. Widely deployed public-key cryptographic schemes are vulnerable to future large-scale quantum attacks, making the migration toward PQC increasingly important. Lattice-based, hash-based, and code-based cryptographic schemes currently represent the most practical candidates for quantum-resistant security across applications such as cryptography, blockchain, software security, and IoT systems, although challenges related to computational overhead, key size, and resource limitations remain unresolved. QKD offers information-theoretic security for communication, but its practical deployment is constrained by infrastructure requirements and communication distance limitations. Furthermore, hybrid quantum-classical approaches have emerged as the dominant direction for applications such as malware analysis and intrusion detection, where shallow quantum circuits combined with classical learning techniques provide a more feasible solution under current NISQ hardware limitations.

Overall, the surveyed literature highlights that quantum computing simultaneously creates new security risks and new opportunities for building advanced cybersecurity solutions. The urgency of the store-now-decrypt-later threat, the heavy dependence on a limited set of quantum hardware platforms for experimental validation, and the gap between theoretical quantum advantage and practical deployment remain important open challenges for the research community. We have tried to provide a comprehensive foundation for researchers and practitioners by consolidating recent developments, identifying existing limitations, and outlining future research directions toward secure, scalable, and trustworthy quantum computing ecosystems.

\section*{Declaration of generative AI and AI-assisted technologies in the manuscript preparation process}

During the preparation of this work, the author(s) used \textbf{GPT-5.5} to assist with manuscript restructuring, thematic reorganization of subsections, and grammatical refinement. After using this tool, the author(s) reviewed and edited the content as needed and take(s) full responsibility for the content of the published article.

\bibliographystyle{elsarticle-num-etal}
\bibliography{GCON}

\end{document}